\documentclass[a4paper,11pt]{article}
\pdfoutput=1 

\usepackage{jheppub} 
                     
\usepackage[dvipsnames]{xcolor}
                     
\newcommand{\be}{\begin{equation}}
\newcommand{\ee}{\end{equation}}

\DeclareMathOperator{\Lvert}{\Bigg\vert}
\DeclareMathOperator{\non}{\nonumber}

\DeclareMathOperator{\OO}{\mathcal{O}}
\DeclareMathOperator{\la}{\langle}
\DeclareMathOperator{\ra}{\rangle}

\title{A modular toolkit for bulk reconstruction}

\author[]{Thomas Faulkner,}
\author[]{Min Li}
\author[]{and Huajia Wang}

\affiliation[a]{Department of Physics, University of Illinois, 1110 W. Green St., Urbana IL 61801-3080, U.S.A.}

\abstract{ 
We introduce new tools for studying modular flow in AdS/CFT. These tools allow us to efficiently extract bulk information related to causality and locality. For example, we discuss the relation between analyticity in modular time and entanglement wedge nesting which can then be used to extract the location of the Ryu-Takayanagi (RT) surface directly from the boundary theory. Probing the RT surface close to the boundary our results reduce to the recent proof of the Quantum Null Energy Condition. We focus on heavy probe operators whose correlation functions are determined by spacelike geodesics. These geodesics interplay with the RT surface via a set of rules that we conjecture and give evidence for using the replica trick. }

\begin{document} 
\maketitle
\flushbottom

\section{Introduction}

A promising approach to quantum gravity comes from a class of dualities where gravity emerges from non-gravitational degrees of freedom that are inherently quantum mechanical yet in principle under control non-perturbatively \cite{Maldacena:1997re,banks1997m}. The quantum information aspects of these degrees of freedom provide a direct way to reveal the emergence of gravity, thus connecting the paradigm of emergence to BH thermodynamics \cite{Bekenstein:1973ur,Hawking:1974sw} and the entropic nature of Einstein's equations \cite{Jacobson:1995ab}. Making these statements precise is an ongoing challenge. Most optimistically we might be able to use an improved understanding of this paradigm to liberate ourselves from the models that are most under control - that of quantum gravity in asymptotically AdS spaces.

Toy models, which are for the most part lattice based quantum systems, that provide a playground for these ideas have been recently developed highlighting the entanglement structure \cite{Ryu:2006bv} and quantum error correcting aspects of holography elucidated in \cite{Almheiri:2014lwa}.
Examples include tensor networks such as MERA \cite{Swingle:2009bg} and holographic error correcting codes \cite{pastawski2015holographic,hayden2016holographic}. 
However including dynamics in these toy models is less clear, and an especially difficult problem involves finding models with emergent locality on the fine-grained or sub-AdS scale.

Starting instead in the continuum limit with the boundary theory as a CFT one automatically builds in dynamics, however dealing with the quantum information aspects of CFTs becomes more tricky.  In the CFT context the usual assumptions for identifying bulk locality \cite{Heemskerk:2009pn} include a large $N$ limit, via factorization rules for correlation functions, and strong interactions, via the requirement of a gap in anomalous dimensions for higher spin single trace operators. Ultimately we would like to combine these assumptions to study the quantum information (QI) aspects of the CFT where bulk emergence should be synergistically linked and such things as the Einstein's equations will naturally and entropically arise. 

One example consequence of these later assumptions is to the chaos bound \cite{Shenker:2013pqa,Maldacena:2015waa} and it's saturation, thus linking strong chaotic behavior (or fast scrambling \cite{sekino2008fast}) to holographic theories with classical gravity duals. The chaos bound, in it's simplest manifestation, constrains the behavior of a CFT four point function continued into real time in a kinematic regime related to high energy Regge scattering in the gravitational dual \cite{shenker2015stringy}. Scattering probes the underlying locality of the bulk \cite{maldacena2017looking,Penedones:2010ue} which has mostly been studied in cases with lots of symmetry (see however \cite{Engelhardt:2016wgb}). In this paper we would like to develop a hybrid approach where we still study a form of bulk scattering yet we also include aspects of quantum information so we can easily work in more general backgrounds and potentially reach behind horizons.

Modular Hamiltonians are basic to the properties of QI in AdS/CFT as first elucidated in the JLMS result \cite{Jafferis:2015del}. Modular Hamiltonian's are associated to the entanglement wedge \cite{Czech:2012bh,Headrick:2014cta,Wall:2012uf} region between the extremal surface that computes the entanglement entropy and the boundary region.  An understanding of the duality of these modular Hamiltonians and it's relation to relative entropy was used to prove that it is possible to reconstruct bulk operators inside the entanglement wedge from the boundary sub-region \cite{Dong:2016eik}.
The ideas behind this proof are the same for the holographic codes as for established holographic theories. However sub-AdS bulk locality is still not clear in the former. 
General modular flow is dual to local boosts deep in the bulk and such boosts might only be expected for theories with sub-AdS locality. We will use such boosts as a tool to extract information on the bulk, and more realistic tensor network models would need to be sensitive to this.

As an example, let us introduce the following correlator, motivated by the recent paper studying the Quantum Null Energy Condition \cite{Balakrishnan:2017bjg}:
\begin{equation}
\label{suchas}
i \mathcal{M} +1 = \frac{\left< \mathcal{O}\Delta^{is}_B \Delta^{-is+1/2}_A \mathcal{O} \right>}{ \sqrt{ \left< \mathcal{O} \Delta^{1/2}_A \mathcal{O} \right>
 \left< \mathcal{O}\Delta^{1/2}_B \mathcal{O} \right>}} 
\end{equation}
where $\Delta_{A,B}$ are modular operators associated to the state $\psi$ reduced to two nested regions satisfying the nesting condition $\mathcal{D}(B) \subset \mathcal{D}(A)$ for their domains of dependence or causal completions. Expectations are taken in the state $\psi$ and $\mathcal{O}$ is a boundary operator in the algebra associated to $\mathcal{D}(B)$. Note that $\mathcal{O}$ should be an appropriately smeared and bounded operator although we will often sweep the details of this under the rug.

Setting $E = i e^{-2\pi s}$ one can show that $\mathcal{M}(E)$ is analytic and bounded in the lower half $E$-plane with $| i \mathcal{M} + 1| \leq 1$ (see Section~\ref{sec:ewn}).
Then following Section~2.3 of \cite{Caron-Huot:2017vep} and using a sum rule, it is not hard to show that:
\begin{align}
\label{anec}
{\rm Im} \mathcal{M} ( E= x - i y) \geq 0 \\ \left|  y \partial_y  \ln {\rm Im} \mathcal{M} ( E =x - i y )\right| \leq 1
\label{modchaos}
\end{align}
for $y>0$.
The later is the so called bound on chaos \cite{Maldacena:2015waa}. For example if we consider $B$ to be a ``small'' deformation of $A$ which we parameterize by some length $\delta z$  then we would like to argue that this correlator looks like:
\be
\mathcal{M} = i p  \, \delta z \,  e^{ \lambda_L s} + \ldots
\ee
where we have also taken somewhat large $s$. The former \eqref{anec} tells us that $p \delta z \geq 0$ and the later bound \eqref{modchaos} tells us that $\lambda_L \leq 2\pi$. We will call this the modular chaos bound. Part of the goal of this paper is to compute the correlator in holographic theories where $\mathcal{O}$ is some boundary local operator not-necessarily close to the boundary entangling cuts at  $\partial B$ and $\partial A$. We will find saturation of the chaos bound $\lambda_L = 2\pi$. In addition to this we find that $\delta z$ is a specific null coordinate shift of the bulk RT/entangling surface associated to the boundary deformation between the two nested regions and that $p$ is a positive quantity related to the operator probe $\mathcal{O}$. 

Thus the emergence of sub-AdS bulk locality in the state $\psi$ should be associated to saturation of the modular Chaos bound $\lambda_L =2\pi$ for a large set of nested boundary regions. We can also use this result to extract information about the location of the entangling surface in the bulk - allowing us to map out the bulk geometry directly from the boundary. Positivity of $\delta z$ turns out to be the nesting requirement of entanglement wedge regions (EWN), given that the boundary regions are nested. 


In order to precisely extract this information we need the correlator to probe localized regions in the bulk and we can achieve this by considering operators that have large conformal dimension $\Delta_{\mathcal{O}}$ which are then dual to large mass $m$ fields in the bulk. Correlation functions are then determined by spacelike geodesics that can be tuned to probe the entangling surface and modular flow. Having access to such a operator is perhaps more than we would like to require of a gravitational dual, although we consider this a mild assumption. This requirement is linked to the large-$N$ limit since the dimension must still be smaller than $N^{\#}$ for any of our discussions to work which will ignore backreaction.  We might employ other brane like objects (one dimensional or more) in the probe limit, such as another entangling surface which is computed via the replica limit of twist operator correlation functions. An entangling surface would eliminate the need for the existence of some other operator, yet the probe would necessarily be co-dimension $2$.  The idea of studying how entangling surfaces themselves are effected by modular flow rather than heavy probe operartors is pursued in a paper that has some overlap with our results \cite{Chen:2018rgz}.\footnote{We thank Aitor Lewkowycz for sharing an early draft of their work, and several discussions leading up to this work.}

The paper is organized as follows. We start by outlining a set of rules for getting a handle on modular flow correlators of heavy probe operators, such as \eqref{suchas}, in AdS/CFT. In Section~\ref{somerules} we simply state the rules, give some intuition for where they come from and check some trivial examples. In Section~\ref{applications} we give some applications, including emphasizing the special role of the operator $\Delta^{1/2}$ which under the rules of Section~\ref{somerules} effectively reflects the boundary operator at the entangling surface. We then study the correlator \eqref{suchas} to extract the bulk EWN condition. In Section~\ref{derivation} we give some more complete arguments for the rules using various forms of the replica trick and related arguments. 

\section{Some rules}
\label{somerules}

For the bulk QFT we can consider a free scalar Klein-Gordan field $\phi$, which is dual to the single trace boundary operator $\mathcal{O}$ with mass $m \ell_{AdS} = \Delta_{\mathcal{O}}$ for large conformal dimension $\Delta_{\mathcal{O}}$. We will demand that the mass is large but not too large as to necessity dealing with back-reaction on the bulk geometry:
$1 \ll m \ell_{AdS} \ll \ell_{AdS}/\ell_p\,, \, \ell_{AdS}/\ell_s$
where $\ell_{p,s}$ are the Planck and string scale respectively. In this limit space-like correlators can be computed using space-like geodesics. This can be shown within the first-quantized approach to computing correlation functions in free QFT - we must fix a Euclidean manifold $\mathcal{B}_E$ and consider maps $x(\tau)$ from the wordline $\mathcal{W}$ to $\mathcal{B}_E$:
\be
\left<  \phi(x) \phi(y) \right>_{\mathcal{B}_E} = 
\int_{x(0) = x}^{ x(1) = y} \frac{\mathcal{D} x(\tau) \mathcal{D} e(\tau) }{V_{\rm diff}} e^{ - S_\mathcal{W} }
\ee
where we take the wordline to be an interval $[0,1]$ parameterized by $\tau$. 
Two point functions are computed by integrating over all such maps into $\mathcal{B}_E$ with the boundary conditions specified above. 
Here $e(\tau)$ is the einbein and dividing by $V_{\rm diff}$ is necessary due to overcounting coming from wordline diffeomorphism invariance. The action is $m$ times the length of the curve:
\be
\label{sw}
S_{\mathcal{W}} = \frac{m}{2} \int d\tau \left( \frac{1}{e} g_{\mu\nu}(x(\tau)) \partial_\tau x^\mu(\tau)  \partial_\tau x^\nu (\tau)
+ e \right)
\ee
which is manifest after integrating over $e$ which sets:
\be
e^{2} = g_{\mu\nu}  \partial_\tau x^\mu(\tau)  \partial_\tau x^\nu(\tau)
\ee

We can interpret this correlation function in a real time/Hilbert space picture in the case where $\mathcal{B}_E$ has a moment of time reflection symmetry such that the Euclidean path integral generates a state $\left| \psi \right> \left< \psi \right|$ at this moment. In which case the world line path integral computes:
\be
\left< \psi \right| \phi(x) \phi(y) \left| \psi \right>
\ee
For sufficiently analytic $\mathcal{B}_E$ we can deform the integration contour of $x(\tau)$ to complex values allowing for the end points $x,y$ to be in real times - that is lying on the real time section of the original space: $x,y \in \mathcal{B}$. The most natural object that we might expect to land on under such a continuation is the time ordered Feynmann propagator.

In the large $m$ limit we can evaluate this in a saddle point approximation as:
\be
\left< \psi \right| \phi(x) \phi(y) \left| \psi \right>
= ({\rm det}_\mathcal{W}) e^{ - m \ell_{\rm cl}(x,y)} \left( 1 + \mathcal{O}(m^{-1}) + \ldots \right)
\ee
where $\ell_{\rm cl}(x,y)$ is the geodesic length between the points $x,y$ and $({\rm det}_{\mathcal{W}})$ is related to the one loop determinant of fluctuations in $x(\tau),e(\tau)$ around the classical saddle. We will not attempt to evaluate this determinant for now, and the semiclassical length will suffice for this paper.

In AdS/CFT we will take the points $x,y$ to the boundary where we then use the extrapolate dictionary to extract the CFT correlation function. This step requires suitable regularization of the of the length of the curve, which has infinite proper length near the boundary of AdS. The procedure also extracts the usual $z^{\Delta_{\mathcal{O}}}$ wavefunction pieces in the dictionary. This is a simple task, the details of which we do not go into. We write the result as:
\be
\left<  \mathcal{O}(x) \mathcal{O}(y) \right>
\approx  e^{ - m \ell_{\rm reg}(x,y)} \rightarrow e^{ - m \ell(x,y)}
\ee
where now $x,y$ lie on the boundary of our assymptotically AdS space and we will often drop the $\rm{reg}$ and simply write $\ell(x,y)$. More specifically the function $\ell(X,Y)$ is defined to be the geodesic length between two bulk points $X,Y$ or the regularized length if one or both of the points are taken to live on the boundary. The geodesic approximation for correlation functions in AdS/CFT has a long history with many applications. See for example: \cite{Balasubramanian:1999zv,Louko:2000tp,Kraus:2002iv,Shenker:2013pqa,engelhardt2014holographic,engelhardt2015further,Maxfield:2017rkn}. In particular we should caution the reader that the geodesic approximation is not fool proof and can sometimes give misleading results, such as acausal response etc.\footnote{We thank Veronika Hubeny and Mukund Rangamani for emphasizing this to us.} These difficulties can often be avoided by appealing to a Euclidean path integral which we will do througout. 

Now that we have setup the computation of correlation functions we would like to introduce modular flow and compute:
\be
\label{modguess}
\left< \mathcal{O}(x) \Delta^{is}_A \mathcal{O}(y)  \right>
\mathop{\approx}^? e^{ - m \ell_s(x,y)}
\ee
for which we might expect that at large $m$ the answer again is determined by some classical solution/geodesic with some associated, but as of now unspecified ``length'' $\ell_s(x,y)$.

At this point we use the JLMS \cite{Jafferis:2015del} result which states that bulk modular flow equals boundary modular flow. Associated to the region $A$ on the boundary there is a bulk region $a$ between $A$ and the associated HRT extremal surface $m_A$ \cite{Ryu:2006bv,hubeny2007covariant}. The quantum fields $\phi$ reduced to this region $a$ define a modular operator $\Delta_a$ that we flow with. The location of the extremal surface $m_A$ plays an important role below which we specify here. This is some co-dimension $2$ surface anchored at the boundary region $\partial A$:
\be
X_A(\xi) : \partial a_A \rightarrow \mathcal{B}_{d,1}
\ee
where $\mathcal{B}_{d,1}$ is the bulk $d+1$ dimensional spacetime with an appropriate conformal boundary. The $\xi^i$ are coordinates on the entangling surface and $X_A^\mu(\xi^i)$ is the mapping into the bulk. The  ($d-1$)  vectors $\Pi_i = \partial_i X_A^\mu \partial_\mu$ are projectors onto the surface.  There are two null orthogonal directions to this surface $k^\pm$ satisfying:
\be\label{1.1.1}
(k_+)^2 = (k_-)^2 = 0 \qquad k_+ \cdot k_- = -1/2 \qquad k_+ \cdot \Pi_i = 0 
\ee
The two complementary entangling wedges are defined as $\mathcal{E}_a = \mathcal{D}(a)$ and $\mathcal{E}_{\bar{a}} = \mathcal{D}(\bar{a})$. 
We will now conjecture some rules, which give situations where we can compute the appropriate $\ell_s(x,y)$ above \eqref{modguess}. We will then discuss a heuristic derivation after we have introduced the rules. In a Section~\ref{derivation} we will give a more complete derivation using the replica trick, but we would like to avoid discussions of replicas for now. We emphasize that these rules do not give a complete solution to modular flow in the large $m$ limit, however they are sufficient to extract lots of interesting bulk physics.

\subsection*{Rule 1}

\begin{figure}[tbp]
\centering 
\includegraphics[width=.48\textwidth]{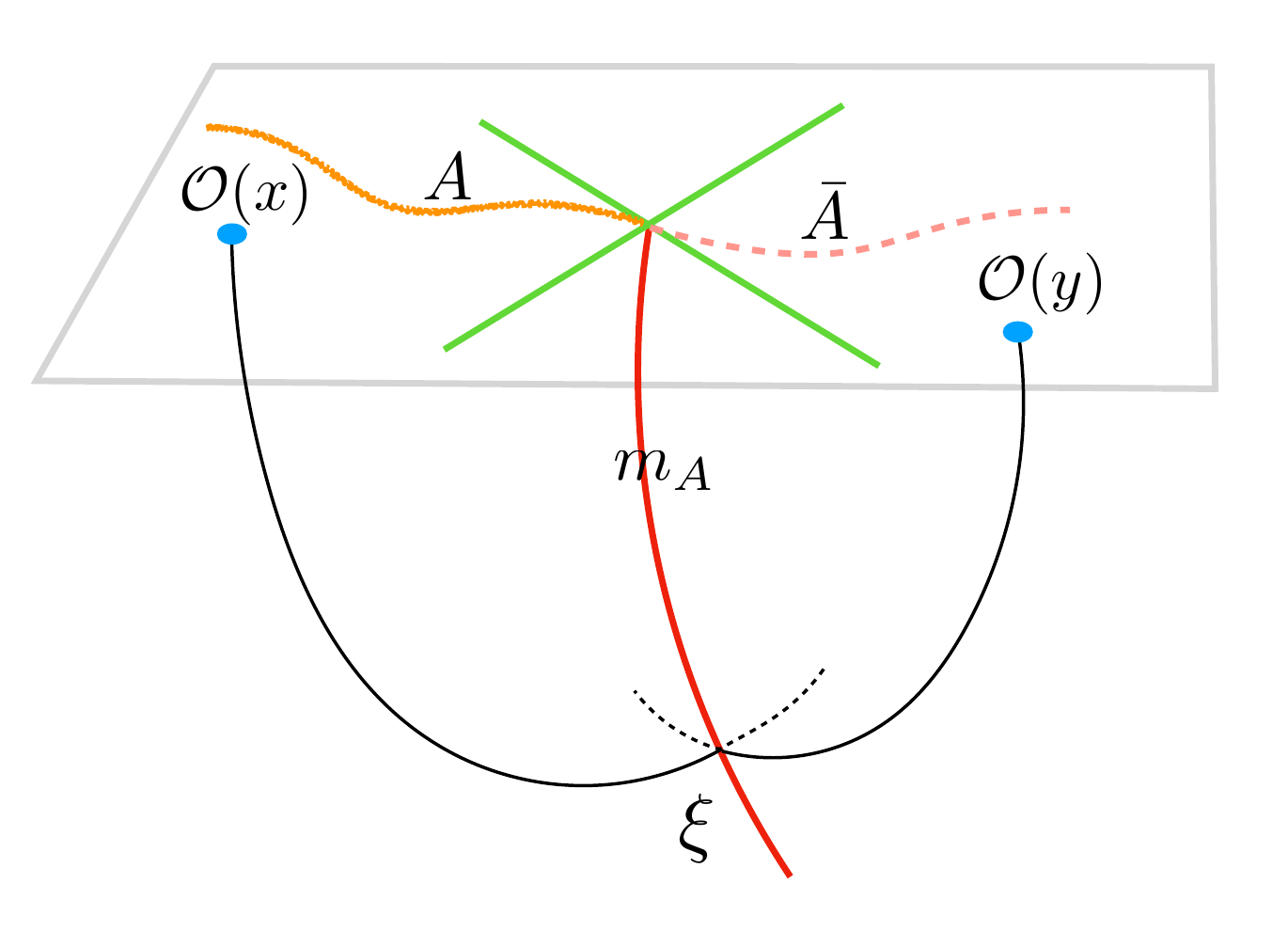}
\hfill
\includegraphics[width=.5\textwidth]{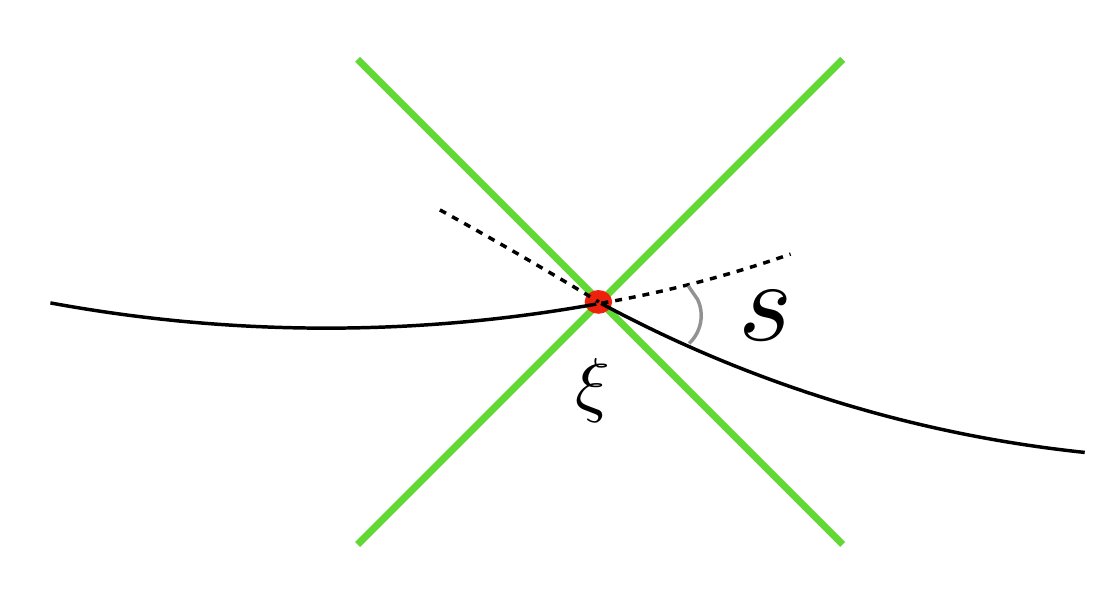}
\caption{\label{fig:rule1} The first rule pertains to a configuration of operator insertions $(x,y)$ and entangling surface $m_A$ such that the smallest length configuration of spacelike geodesics that meet at some point  $\xi$ on the entangling surface with a local relative boost of rapidity $2\pi s$. The modular flow correlator is then computable as a sum of lengths. The right figure is a zoom of the left figure showing only the transverse directions to $m_A$.  }
\end{figure}

Consider the  piecewise geodesic from the boundary $x$ in $\mathcal{D}(A)$ to $\xi$ which is some location on $m_A$ and back up to the boundary at $y$ now in $\mathcal{D}(\bar{A})$ . If the two segments of the piecewise geodesic are locally related via a boost around the location $X_A(\xi)$ (see Figure~\ref{fig:rule1}) with rapidity $2\pi s$ then:
\be
\left<  \mathcal{O}(x) \Delta_A^{is} \mathcal{O}(y)\right>
\approx  \exp\left( - m \left[ \ell(x,X_A(\xi)) + \ell(X_A(\xi),y)\right] \right)
\ee
where the appropriate boundary regularization of the length is implied. We can construct a piecewise curve $x(\tau)$ which includes both segments of the geodesic joined together at $\tau=\tau_m$ with $0< \tau_m < 1$ and such that $x(\tau_m \pm \epsilon ) = X_A(\xi)$. The local boost condition can be written out explicitly using:
\begin{equation}
\label{defn}
n_i(\tau) = e^{-1} \partial_\tau x(\tau) \cdot \Pi_i \qquad n_\pm(\tau) = e^{-1} \partial_\tau x(\tau) \cdot k_\pm
\end{equation}
such that:
\begin{align}
\label{match1}
(n_i)_{\partial a} &= (n_i )_{\partial \bar{a}}  \\
\label{match2}
(n_+,n_-)_{\partial a} &= (e^{-2\pi s} n_+, e^{2\pi s} n_-)_{\partial \bar{a}} 
\end{align}
where the matching occurs approaching the extremal surface from either of the  two entanglement wedges.  The notation is such that $\partial_a : \tau= \tau_m - \epsilon$ and $\partial_{\bar{a}}: \tau= \tau_m + \epsilon$.


Note that this will only work for a co-dimension one slice of the $2d+1$ parameters $\{x,y;s\}$ specified in the correlator.  One way to find such a set $\{x,y;s\}$ is to leave unspecified $s$ and minimize over the point $\xi$ on the surface $m_A$. This minimization procedure then guarantees that the $i$ components of \eqref{match1} match smoothly.
After which we can compute $s$ as:
\be
\label{sxy}
s = s(x,y) \equiv \frac{1}{4\pi} \ln \left[ \left( \frac{ \partial_\tau x(\tau_m) \cdot k_+}{ \partial_\tau x(\tau_m) \cdot k_-}\right)_{\partial a} \left(\frac{  \partial_\tau x(\tau_m) \cdot k_-}{ \partial_\tau x(\tau_m) \cdot k_+}\right)_{\partial \bar{a}} \right]
\ee
such that the other components of \eqref{match2} are satisfied. Since $\xi(x,y)$ are unspecified parameters that are fixed by the minimization procedure it is clear that we have determined a $2d$ slice of parameter space, which we might write $\{x,y;s(x,y)\}$. 

\subsection*{Rule 2}

\begin{figure}[tbp]
\centering 
\includegraphics[width=.48\textwidth]{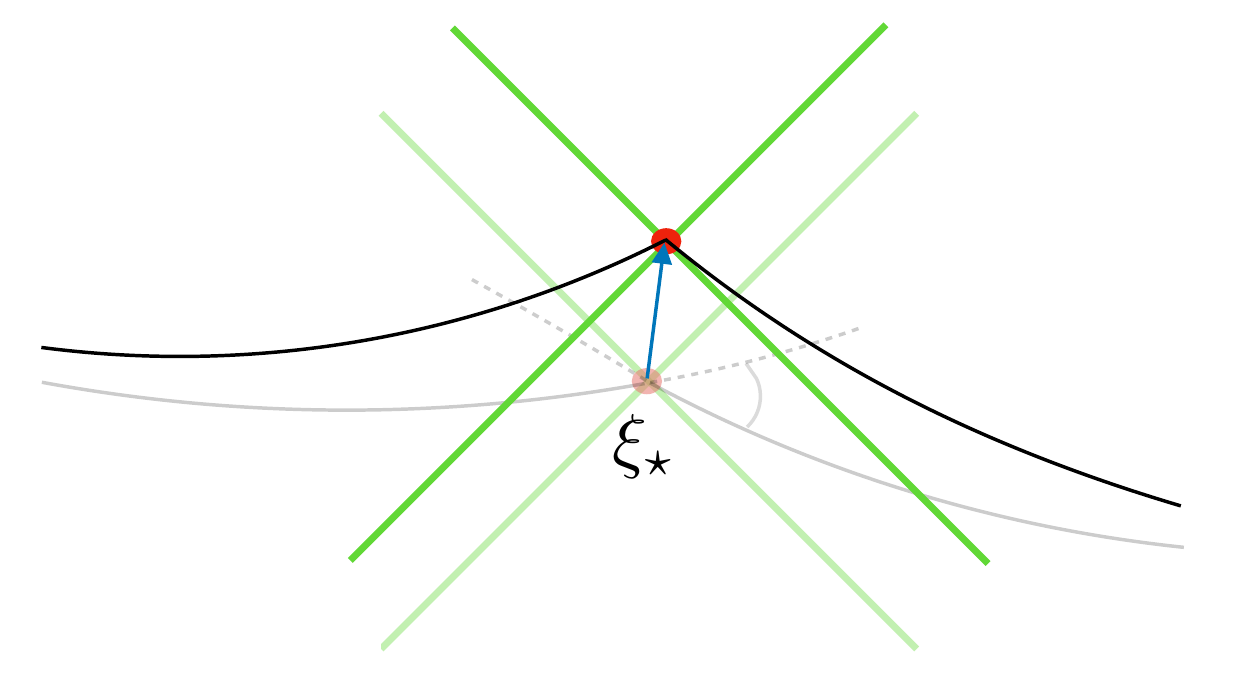}
\hfill
\includegraphics[width=.5\textwidth]{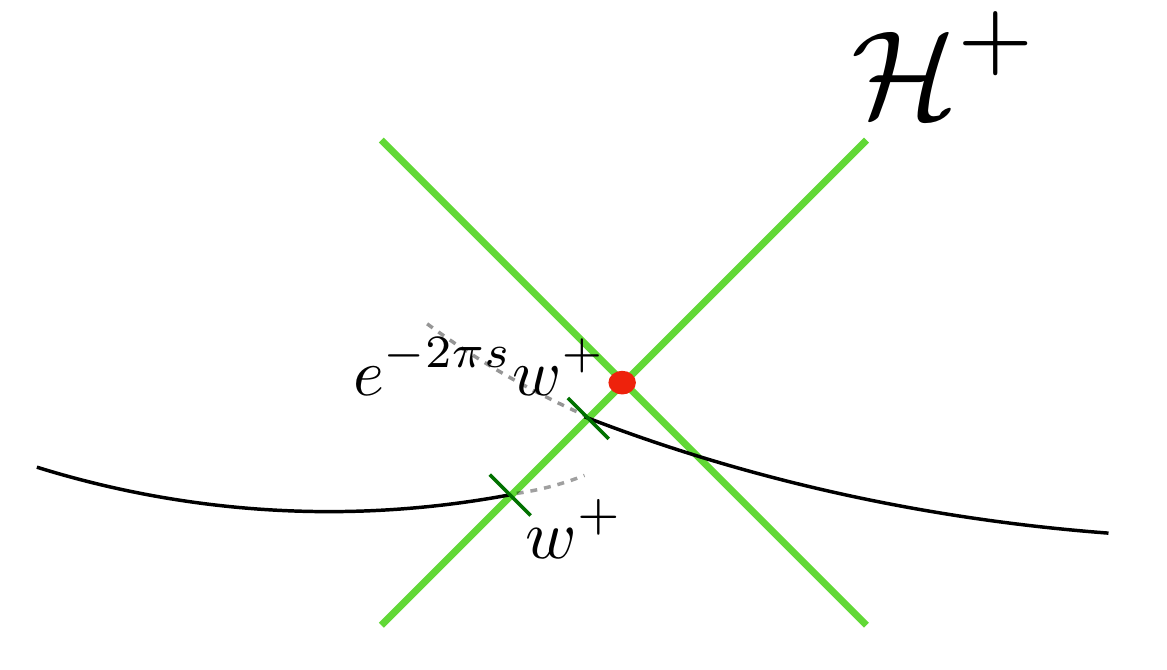}
\caption{\label{fig:rule2} The second rule pertains to a small deformation of the picture in \ref{fig:rule1}. In other words first order deformations in the parameters that lead to computable modular flow are still computable to second order in those deformations using a set of spacelike geodesics that meets on the actual entangling surface. The right figure shows an alternative description which is useful for calculations. Note that a choice to match in a related way along $\mathcal{H}^-$ would lead to the same result.  }
\end{figure}

The second rule pertains to a small deformation of the above rule. Consider the set of parameters $\{x,y;s_\star \}$ and entangling cut $\partial A$ which satisfy Rule 1 with the geodesic intersecting
$m_A$ at $\xi_\star$. If we deform either $s$ or the location of the entangling cut on the boundary $A \rightarrow B$ then we claim that up to and including first order in these deformations we can still compute the leading classical contribution. The answer is:
\begin{align}
\label{rule2}
\left<  \mathcal{O}(x) \Delta_B^{is} \mathcal{O}(y) \right>
\approx & \exp\left( - m \left[ \ell(x,X_B(\xi)) + \ell(X_B(\xi),y) 
+ \mathcal{O}(\delta^2) \right] \right)
\end{align}
where $\delta$ includes either $\delta s = (s-s_\star)$ or $\delta z^\pm$ where the later are the resulting small deformations, in the null normal directions, of the RT entangling surface at the point $\xi_\star$. These deformations are a result of deforming the boundary region $\partial A \rightarrow  \partial B$. 

We note that there is an ambiguity in moving the coordinate from the initial location $\xi_\star$ on $m_{A}$ (satisfying the conditions in Rule 1) to the point $\xi$ on $m_B$. As long as we transport the coordinates as we move the surface in a smooth continuous fashion then this ambiguity will only effect the $\mathcal{O}(\epsilon)^2$ correction and so we may safely ignore this issue. This is because of the momentum conservation of the undeformed geodesics  transverse to the surface \eqref{match1}. To make a slightly more specific choice we will transport those coordinates that are along the surface $m_{A}$, perpendicular to the surface. This natural choice can be specified in Gaussian normal coordinates: 
\be
\label{gn}
ds^2 = - d w^+ d w^- +  
 h_{ij} d \xi^i d \xi^j
+ \mathcal{O}(w) 
\ee
where $\alpha,\beta = \pm$ are the null coordinates perpendicular to the surface. The original entangling surface $m_{A}$ lives at $w^\pm =0$
 and the deformed surface lives at $B: w^\pm = \delta z^\pm(\xi)$. 
If the geodesic originally intersects $m_{A}$ at some $\xi_\star^i$ then we simply take the point on $B$ to be at $(\xi_\star , w^\pm = \delta z^\pm(\xi_\star))$. 

We can give a slightly different version of rule $2$ that is convenient for calculations later. Consider,
close to the entangling surface, the null surfaces $\partial \left( \mathcal{E}_b \cup  \mathcal{E}_{\bar b} \right) = \mathcal{H}^+ \cup \mathcal{H}^-$. These can be defined as the interesecting light-sheets shot out from $m_B$ to the past and future. In the coordinates of \eqref{gn} they sit at $w^+ =0$ or $w^-=0$.

Take one of these surfaces, say $\mathcal{H}^+$, and pick some point $Z \in \mathcal{H}^+$ which is assumed to be close to the entangling surface $m_B$. We then construct a piecewise continuous geodesic that passes from the boundary at $x$ to this null surface at $Z$ and then from the boosted point $e^{-2\pi s \chi} (Z)$ on the null surface back to the boundary at $y$. Here $\chi$ is a vector field that is locally a boost at $m_B$ but otherwise away from the surface it is arbitrarily specified. In the coordinates of \eqref{gn} we can pick this to be:
\be
\label{chi}
\chi = w^+ \partial_+ - w^- \partial_-
\ee
We claim that extremizing over $Z$ determines the correlator up to $\mathcal{O}(w^{+})^2$ corrections in the length:
\be
\label{rule2disc}
\left<  \mathcal{O}(x) \Delta_B^{is} \mathcal{O}(y) \right> \approx  \mathop{{\rm ext}}_{Z \in \mathcal{H}^+} \exp \left( - m  \left( \ell(x, Z )+ \ell(e^{-2\pi s \chi}(Z), y)\right)  \right)
\ee

This can be seen to be equivalent to Rule 2 as follows. Working in Guassian normal coordinates
we have $Z = (w^+,\xi^i)$ such that $e^{-2\pi s\chi}(Z)=(e^{-2\pi s}w^+, \xi^i)$. Assuming the geodesics pass close to the entangling surface at $\xi_\star$ then we can approximate the length by varying the worldline action in \eqref{sw} about the solution that does go to $\xi_\star$:
\be
\delta S_\mathcal{W} = \frac{m}{e} \delta x^\nu g_{\mu\nu} \partial_\tau x^\mu
\ee
Applying this to both segments of the geodesic we can calculate the approximate length of the total geodesic:
\begin{align}
\label{vary}
&\ell(x, X_B(\xi_\star) ) + \ell(X_B(\xi_\star),y)  + w^+ \left( (n_+)_{\partial b} - e^{-2\pi s} (n_+ )_{\partial \bar{b}} \right)  \\ & \qquad \qquad \qquad \qquad \qquad
+  (\xi^i-\xi^i_\star) \left( (n_i)_{\partial b} - (n_i)_{\partial{\bar{b}} } \right)  + \mathcal{O}(\delta^2) \nonumber
\end{align}
where $\delta$ is one of either  $w^+$ or  $(\xi^i-\xi^i_\star)$. Including these quadratic order terms,
which we assume generically do not vanish, we can extremize over $w^+$ and $\xi^i$. We find the conditions in (\ref{match1}-\ref{match2}) are approximately satisfied up to $\mathcal{O}(\delta)$ corrections and the length remains the same as the length of the piecewise geodesic passing to the entangling surface up to quadratic corrections $\mathcal{O}(\delta^2)$ in these parameters. So we have linear errors in Rule 1 leading to quadratic errors in the length. This is exactly the content of Rule 2.


\subsection{Some intuition}

\begin{figure}[h!]
\centering 
\includegraphics[width=1\textwidth]{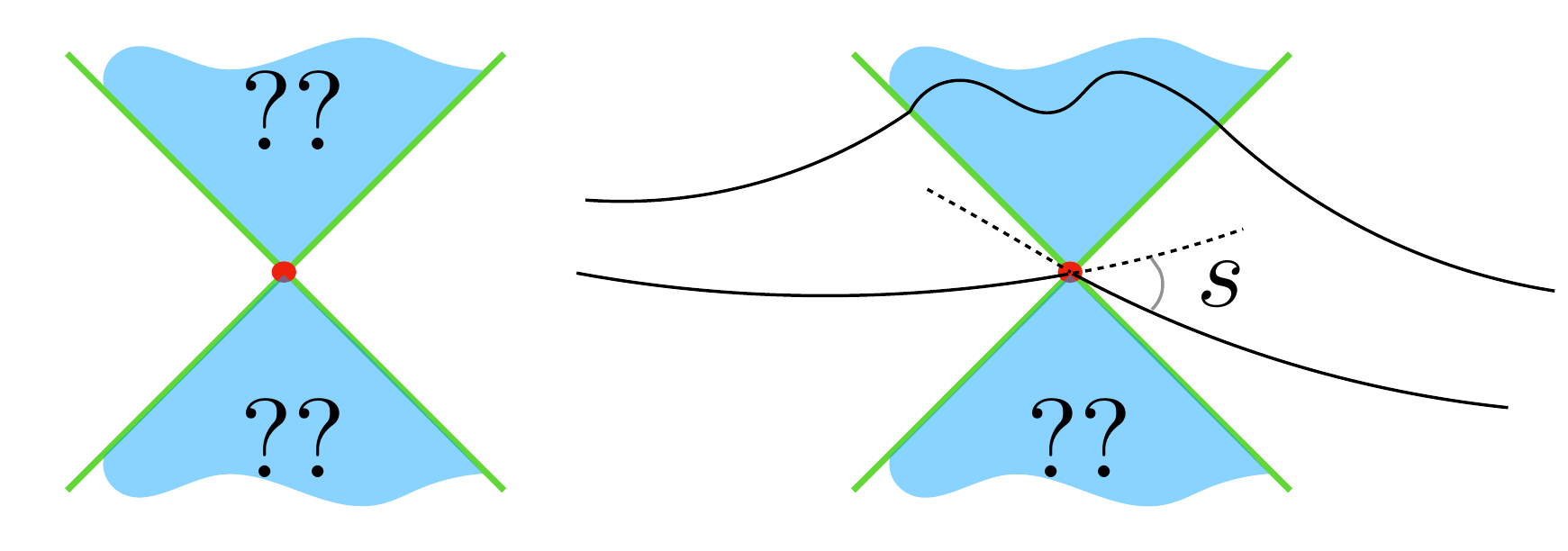}
\caption{ \label{showfig} Modular flowed geometry. The white regions are the two entanglement wedges that are the same as in the unflowed state. The blue region is effected by modular flow and in general unknown. In the right section we show some sample geodesics that compute correlation functions. The geodesic may enter the unknown region in which case we do not have control. If we manage to tune parameters so that this does not happen, we are left with a simple local analysis at the RT surface. }
\end{figure}

In this section we will develop some intuition for the above rules. We will leave a more complete discussion to later in Section~\ref{derivation}. Following \cite{Jafferis:2014lza} we consider the ``modular flowed'' state:
\be
\left| \psi_s \right> = \rho_A^{is} \left| \psi \right>
\ee
where we are explicitly flowing with only half the modular Hamiltonian and not the full $\Delta_A = \rho_A \otimes \rho_{\bar{A}}^{-1}$ which by definition leaves invariant the defining state. Using this later fact we are free to move the flow to the $\bar{A}$ region, $\left| \psi_s \right> = \rho_{\bar{A}}^{is} \left| \psi \right>$.  In this new state we can compute the correlation function of two operators in complementary wedges:
\be
\label{corrflow}
\left< \psi_s \right| \mathcal{O}(x)  \mathcal{O}(y) \left| \psi_s \right>
 =  \left< \psi \right| \mathcal{O}(x) \Delta_A^{is}  \mathcal{O}(y) \left| \psi \right>
\ee
This is the desired correlator. So we simply need to understand how to compute two point functions in the flowed state $\left| \psi_s \right>$. 
This state has the interesting property that it leaves invariant the expectation value of operators completely localized in either $\mathcal{D}(A)$ or  $\mathcal{D}(\bar{A})$:
\be
\label{Taa}
\left< \psi_s \right| T_A \left| \psi_s \right> = \left< \psi \right| T_A \left| \psi \right>
\quad {\rm and} \quad \left< \psi_s \right| T_{\bar{A}} \left| \psi_s \right> = \left< \psi \right| T_{\bar{A}} \left| \psi \right>
\ee
This includes correlation functions of operators all in the same region.
In holography if we were to try to reconstruct the gravitational dual of $\left| \psi_s \right>$, assuming that it can described by some semi-classical geometry $g_s$,
then we would come up with a picture where the description in the entanglement wedges $\mathcal{E}_a$ and $\mathcal{E}_{\bar{a}}$ is completely unaltered. In particular, if we use the ideas of entanglement wedge bulk reconstruction - then the operators in \eqref{Taa} are sufficient to reconstruct the entanglement wedges and so they are left invariant by modular flow.  Thus we can sketch a picture of the bulk spacetime shown in Figure~\ref{showfig} where the shaded regions are the only regions that can be effected by modular flow. These regions are timelike to the past or future of $m_A$. In the local Rindler-like cases they are the Milne wedges and so we will refer to them as such even in the more general case. 

As was argued in \cite{Jafferis:2014lza} and later extended in \cite{Jafferis:2015del}, the shaded regions should be constructed in the bulk by acting with a combination of the area operator and the bulk modular Hamtilonian for the quantum fields in the bulk:
\be
\left| \psi_s \right> = e^{ -i 2\pi s \widehat{A}_{m_A}/4 G_N } \rho_{a}^{is} \left| \psi \right>
\ee
The area operator lives on the entangling surface and is in the center of the bulk operator algebras for the entanglement wedges. There is a non-trivial action to the future of the entangling surface. The combined action of the area operator and the bulk modular flow leaves a non-singular state for the bulk quantum fields \cite{Jafferis:2015del}. This is the case for bulk regions that are at least spacelike from the \emph{boundary} entangling surface $\partial A$.\footnote{The boundary state $\left| \psi_s \right>$ is singular to the bulk future of the \emph{boundary} region $\partial A$ since the half modular Hamiltonian is singular here, although this singularity is not important for computing two sided correlators as shown in \eqref{corrflow}. The geodesics we will use will never enter the singular bulk region to the bulk future/past of $\partial A$.}
We will  assume that such a spacetime can be treated semi-classically with some metric $g_s$ including inside the Milne wedges. 

Following \cite{Jafferis:2014lza} we expect  that in the large $m$ limit the modular flow correlator is computed by an extremal geodesic in this new spacetime $g_s$. Since we will not attempt to construct $g_s$ generally, we will not be able to say much about general modular flow. However if we tune the parameters, such as the locations $x,y$ of the operators on the boundary, to satisfy Rule 1 then we find a situation where the geodesic \emph{never enters into the unknown regions} of $g_s$. We must only understand the matching condition between the left and right wedges. It is not hard to guess that the matching condition should be a local boost by rapidity $2\pi s$. More specifically one can argue that since the area operator is the Noether charge/generator of a diffeomorphism acting like a boost on one side of the entangling surface (where the diffeomorphism equals the vector $\chi$ defined in \eqref{chi}), then the geodesic should match only after an action of this half-sided diffeomorphism. This is obviously true in the local case and it is not hard to imagine that it works the same in the non-local case.

Then Rule 2 can be justified as follows. The relevant curve that computes the approximate length is still anchored to the entangling surface, and even though it does not satisfy the requirements of Rule 1, it is perturbatively close to the correct geodesic in the spacetime $g_s$. This actual geodesic necessarily passes into the Milne wedge so we do not have control over it's length, except to say that since it is a geodesic and thus extremal,  linear deformations only change the length to second order. Since the anchored curve is a linear deformation of the correct geodesic the result of \eqref{rule2} follows. 

There are some unresolved issues in this approach which make this discussion less precise. For example we have not carefully taken into account the effect of the non-trivial state for the bulk quantum fields of $\phi$ on the correlator. While this state is non-singular near the entangling surface, in the non-local case we expect some non-vacuum squeezed state since the modular operator is bi-linear in the quantum fields.
Likely in the large mass limit we can neglect such state dependent effects in the geodesic approximation but we do not have a precise argument.  Indeed from the first quantized/worldline point of view, where the geodesic approximation finds a natural home, it is not obvious how to include properties of the state at all. The best approach is to consider the first quantized path integral where the target is some Euclidean space which then gives a natural state interpretation upon Wick rotation. Following this logic for the modular flow computation leads us to considering the worldline path integral on the Euclidean replicated geometry. This is the approach we will take in Section~\ref{derivation}.

\subsection{The local/Rindler case}

Let us consider these rules in the light of local modular Hamiltonians. The rules are somewhat trivial here. For example let us consider $AdS_3$ in Poincare coordinates:
\be
ds^2 = \frac{d z^2  - d u d v}{z^2}
\ee
and consider the $A$ entangling surface $u = v =0$. In $AdS_3$ geodesics are semi-circles that come in perpendicular to the boundary. Given an operator at location $x = (u_x,v_x)$ in the left wedge (with $u_x >0, v_x <0$) there is a mirror location $x_m = (-u_x, - v_x)$. The correlator of $\left< \mathcal{O}(x) \mathcal{O}(\lambda x_m) \right>$ with $0 < \lambda <\infty$ is computed by a geodesic that clearly passes through the entangling surface at $z = z_0 = \sqrt{ - \lambda u_x v_x}$.
For fixed $x$ then the length of the Rule-$1$  geodesic will always be computed by one of these geodesics. For a given $y$ we should then compute $s$ and $\lambda$ by boosting $\lambda x_m$ to this point:
\be
-\lambda u_x e^{2\pi s} = u_y \qquad -\lambda v_x e^{-2\pi s} = v_y
\ee
such that $\lambda = \sqrt{ u_y v_y/ (u_x v_x)} $ then and in particular 
\be
s(x,y) = \frac{1}{4\pi} \ln \frac{u_y v_x }{u_x v_y} \qquad z_0(x,y) =  ( u_x v_x u_y v_y)^{1/4}
\ee
We can then check that the left and right hand side of:
\be
\label{leftright}
\left< \mathcal{O}(x) \Delta_A^{is(x,y)} \mathcal{O}(y) \right>
= \exp\left( - m \ell(x, z_0 )- m \ell(y,z_0)\right)
\ee
agree. Here $ l(x,z_0)$
is the regularized length of a geodesic between $x$ and the point $z=z_0$ on the entangling surface.
\begin{eqnarray}\label{2.2.1}
\ell[x,z_0]=\ln\bigg(z_0\big(1-\frac{u_x v_x}{z_0^2}\big)\bigg)
\end{eqnarray}
All together we have:
\be
\label{len1}
 \ell(x, z_0 )+ \ell(y,z_0) = \ln\bigg(z_0^2\big(1-\frac{u_x v_x}{z_0^2}\big)\big(1-\frac{u_y v_y}{z_0^2}\big)\bigg) 
\ee
This should be compared to the left hand side \eqref{leftright} which we can compute using boosts on the boundary of the CFT two point function:
\begin{eqnarray}
\la \OO(u_x,v_x)\OO(u_ye^{-2\pi s},v_y e^{2\pi s})\ra=\frac{c_{\Delta}}{\big(-(u_x-u_ye^{-2\pi s})(v_x-v_y e^{2\pi s})\big)^{\Delta_{\OO}}}
\end{eqnarray}
which gives a length of,
\be
\label{len2}
\ln \left( (e^{-2\pi s} u_y - u_x)  (v_x - e^{2\pi s} v_y)  \right)
\ee
Plugging in the specific value of $s=s(x,y)$ and $z_0(x,y)$ we can show that \eqref{len1} agrees with \eqref{len2}. 

Slightly less trivial is the small deformation case. Let us imagine moving the entangling surface in the null direction to $u= \delta z^-$. To leading order we can take the geodesic to go to the point $z_0$ on the entangling surface. The new length of the discontinuous geodesics is:
\be
\ln  z_0 \left( 1 - \frac{(u_x - \delta z^-) v_x}{z_0^2} \right) + \ln  z_0 \left( 1 - \frac{(u_y - \delta z^-) v_y}{z_0^2} \right)
\ee
Expanding we find the correction to the length is:
\be
\label{smagree2}
\delta \ell = \frac{ \delta z^- v_x}{ z_0^2 - v_x u_x} + \frac{ \delta z^- v_y}{ z_0^2 - v_y u_y} +\OO(\delta z^{-2})
\ee
We should compare this correction to the direct calculation of the correlator using the geometric boost. 
Since the entangling surface is slightly shifted the boost now acts on the $u_y$ coordinate as:
\be
u_y \rightarrow ( u_y - \delta z^- ) e^{ -2 \pi s} + \delta z^-
\ee
From which we can compute the correction to the correlator
\be
- m \delta \ell \approx \delta z^- ( e^{-2\pi s} -1 ) \partial_{u_y} \ln \left< \mathcal{O}(x) \Delta_A^{is} \mathcal{O}(y) \right>
\ee
such that:
\be
\label{smagree1}
\delta \ell = \frac{ \delta z^- (e^{-2\pi s} -1)}{ u_x - e^{-2\pi s} u_y} 
\ee
Fixing the boost $s(x,y)$ and $z_0(x,y)$ as specified above one can show that these two expressions \eqref{smagree1} and \eqref{smagree2} agree.

\section{Applications}
\label{applications}

The two rules we outlined are potentially very powerful despite not giving a general picture of modular flow. Dealing with the two operator locations $x,y$, however is cumbersome. Especially if we want to demand that the geodesic probe a fixed point $\xi$ on the entangling surface as we change $s$. This necessitates tuning these insertion points in some way as we vary $s$.  This could be done and results in a boundary curve,  for which:
\be
\label{bdrcurve}
\left< \mathcal{O}(x) \Delta^{is} \mathcal{O}(y_{(x,\xi)}(s)) \right>
\ee
is computable.  A similar boundary curve to $y_{(x,\xi)}(s)$ was discussed recently in the local case \cite{Wen:2018whg}. 
More generally it is simply constructed by shooting out a geodesic from the bulk point $X_A(\xi)$ on $m_A$ in a direction that is the same as the direction of the geodesic that passes from $x$ to $X_A(\xi)$ up to a local boost by an amount $s$ about the entangling surface (that is using the rule \eqref{match2}). The intersection point of this geodesic at the boundary defines the curve $y_{(x,\xi)}(s)$. Perhaps the more natural problem was formulated already around \eqref{sxy} where we hold fixed $x,y$ and allow $\xi$ to vary. They are related via:
\be
y_{(x,\xi(x,y))}(s(x,y)) = y
\ee
There is considerable freedom here and it is not clear what is the best approach.  To remove some of the freedom we will consider a slightly different setup where we take $x,y$ to be in the same boundary wedge region. We turn to this now. 

\subsection{Mirrors for mirror operators}

In order to satisfy the local boost condition for two operators inserted in the same wedge $x,y \in \mathcal{D}(A)$ we must allow for modular flow which is complex. Imaginary boosts local to the entangling surface become Euclidean rotations and the value $s= i/2+t$ rotates the curve by an angle $\pi$ and then boosts it by an amount $t$ such that the geodesic folds back into the same wedge after bouncing off $m_A$. In this case we do not have the intuitive picture involving geodesics in some deformed spacetime $g_s$ to justify this rotated geodesic, however analyticity in $s$ suggests there is no other option and indeed we will give a derivation that allows for complex boosts later. This derivation works particularly well for the case $s = i/2$ which we discuss now.

When $s=i/2$ then the two boundary insertion points must be the same $y=x$ in order to satisfy the local boost condition since the geodesic simply gets reflected back on itself. In other words the function \eqref{sxy} $s(x,y) \rightarrow i/2$ as $x \rightarrow y$ and the boundary curve appearing in \eqref{bdrcurve} continued into complex $s$ must satisfy $y_{(x,\xi)}(i/2) = x$ where $\xi = \xi(x,x)$. This allows us to drop the two different operator locations. Indeed the computable correlator in this case is simply:
\be
\left< \mathcal{O}(x) \Delta^{1/2}_A \mathcal{O}(x) \right> =
\left< \mathcal{O}(x) \mathcal{O}(x)_J \right>  \geq 0
\ee
where we have taken $\mathcal{O}(x)$ to be a Hermitian operator.
So the correlator we study involves the Tomita-Takesaki modular conjugated operator $\mathcal{O}(x)_J = J_A \mathcal{O}(x) J_A \in \mathcal{A}_{\bar{A}} $. Similar operators to $\mathcal{O}(x)_J$ were called mirror operators in \cite{Papadodimas:2013jku}  and here we see that their correlation functions are computed by literally thinking of the entangling surface as a mirror.

\begin{figure}[h!]
\centering 
\includegraphics[width=.45\textwidth]{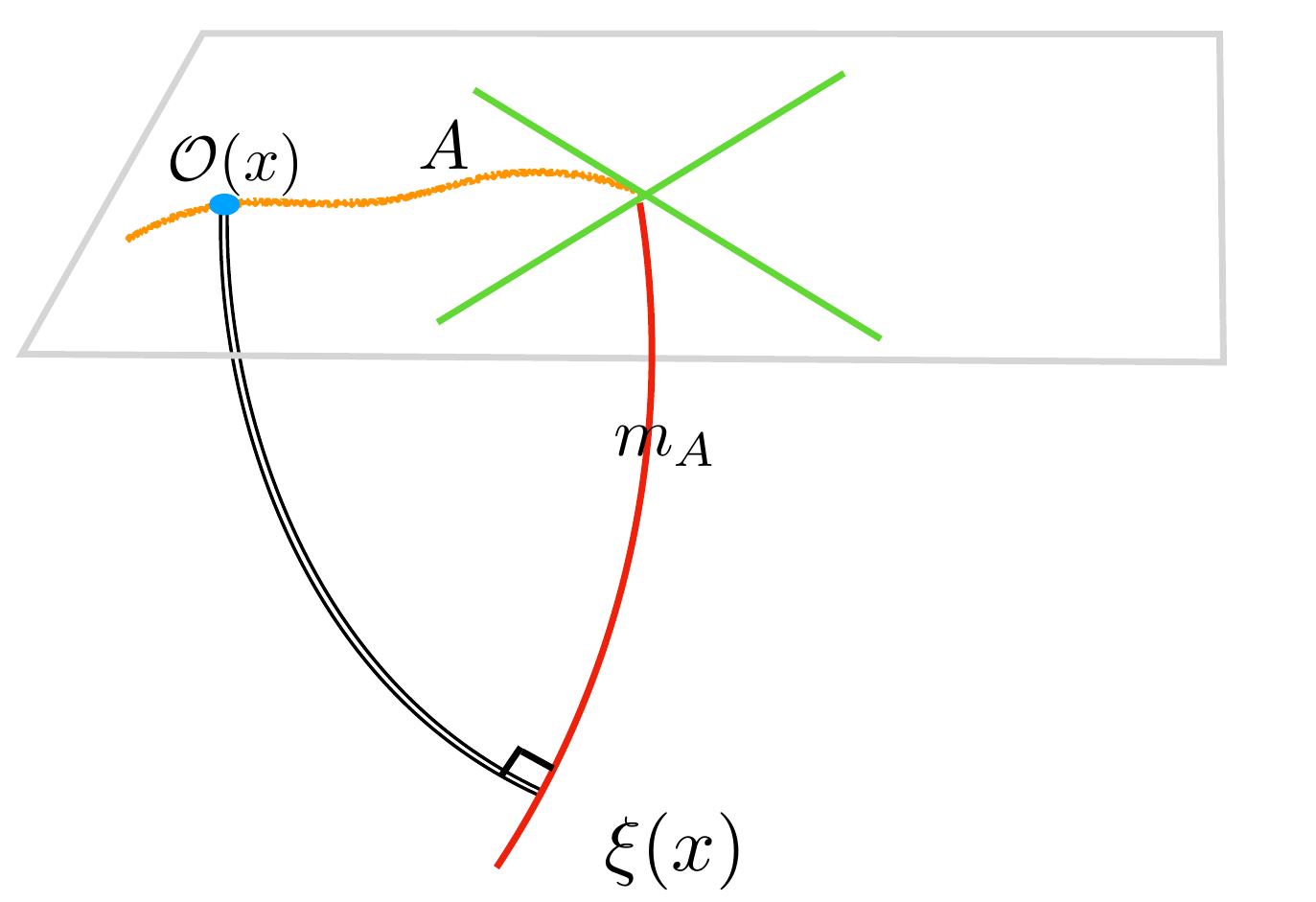}
\hfill
\includegraphics[width=.52\textwidth]{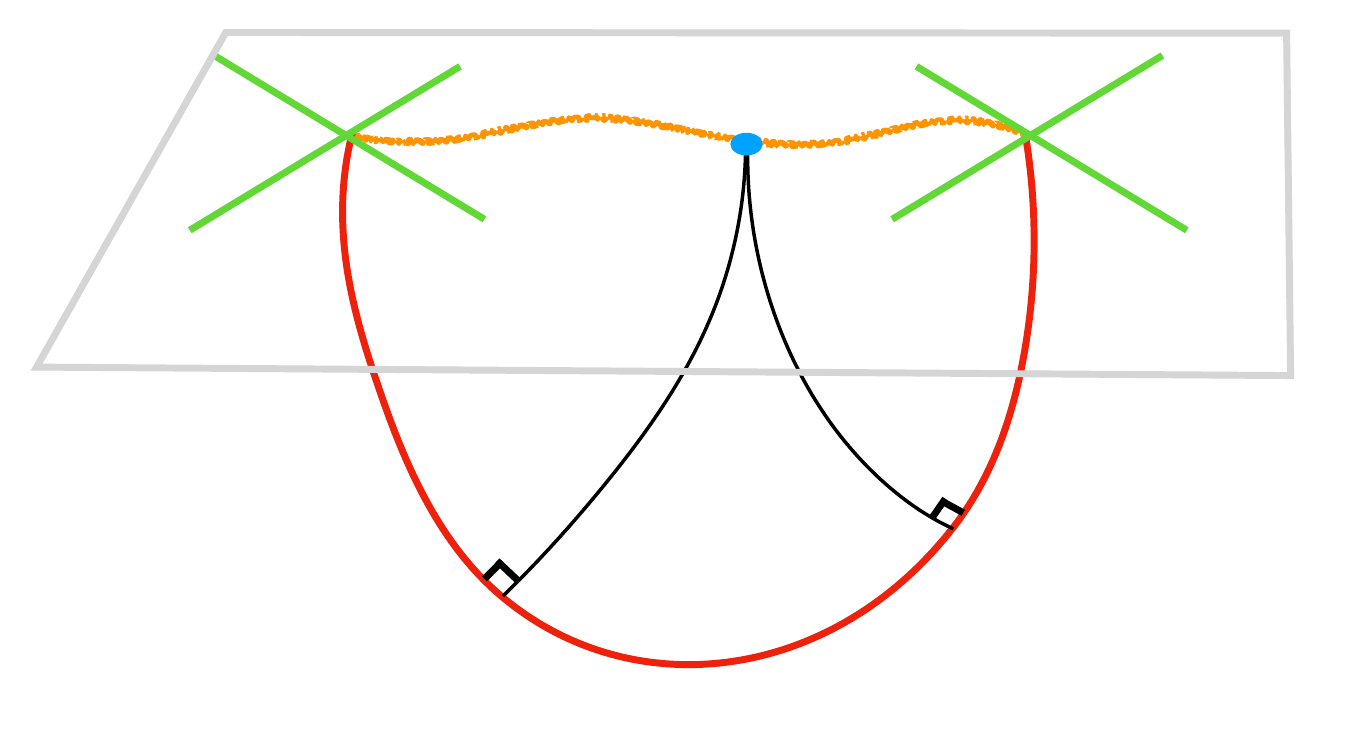}
\caption{Left: Correlation functions of operators with their mirrors, according to the rules, are computed
via a geodesic that reflects of the entangling surface perpendicular to the surface. Right: this map
is not one to one and there might be several points on the entangling surface that are represented by the same point on the boundary. This results in shadows, which are regions on the entangling surface that are not accessible with these mirror correlators. In the above example the shadow lies between the two reflection points. The above example is a cartoon and we have not worked through any real example, leaving this to future work. \label{shadow} }
\end{figure}

As we vary $x$ we map out some coordinate on the entangling surface $\xi(x)
\equiv \xi(x,x)$. We find this point by minimizing:
\be
\xi(x) :\ell (x,X_A(\xi(x))) =  {\rm min}_{\xi'} \ell (x,X_A(\xi'))
\ee
The minimization procedure guarantees that the geodesic hits the entangling surface perpendicular to the surface with $ (e^{-1} \partial_\tau x \cdot \Pi_i)_{\partial a} = 0$ such that a $\pi$-Euclidean rotation then sends the geodesic back on itself with $\partial_\tau x \rightarrow  - \partial_\tau x$ at $\xi$. The correlator is then:
\be
\label{mirrorcorr}
\left< \mathcal{O}(x) \Delta^{1/2}_A \mathcal{O}(x) \right>  \approx \exp\left( - 2 m \ell(x,X_A(\xi(x))) \right)
\ee
The map $\xi(x)$ is from $d$ parameters of $x\in \mathcal{D}(A)$ to $d-1$ coordinates on the entangling surface. Fixing a space-like slice, say the region $x\in A$, on the boundary gives a natural coordinate chart on the entangling surface. These coordinates however will likely breakdown in many cases due to caustics on the space-like congruence of geodesics shot out perpendicular from the entangling surface. In particular $\xi(A)$ will not be the entire entangling surface. There is some form of entanglement shadow - an example of this is sketched in Figure~\ref{shadow}. We will not attempt to overcome this shortcoming, however we note that with more work and using the freedom to tune $x,y$ separately one can get information everywhere on the entangling surface.

\subsection{Entanglement Wedge Nesting}
\label{sec:ewn}

\begin{figure}[h!]
\centering 
\includegraphics[scale=.65]{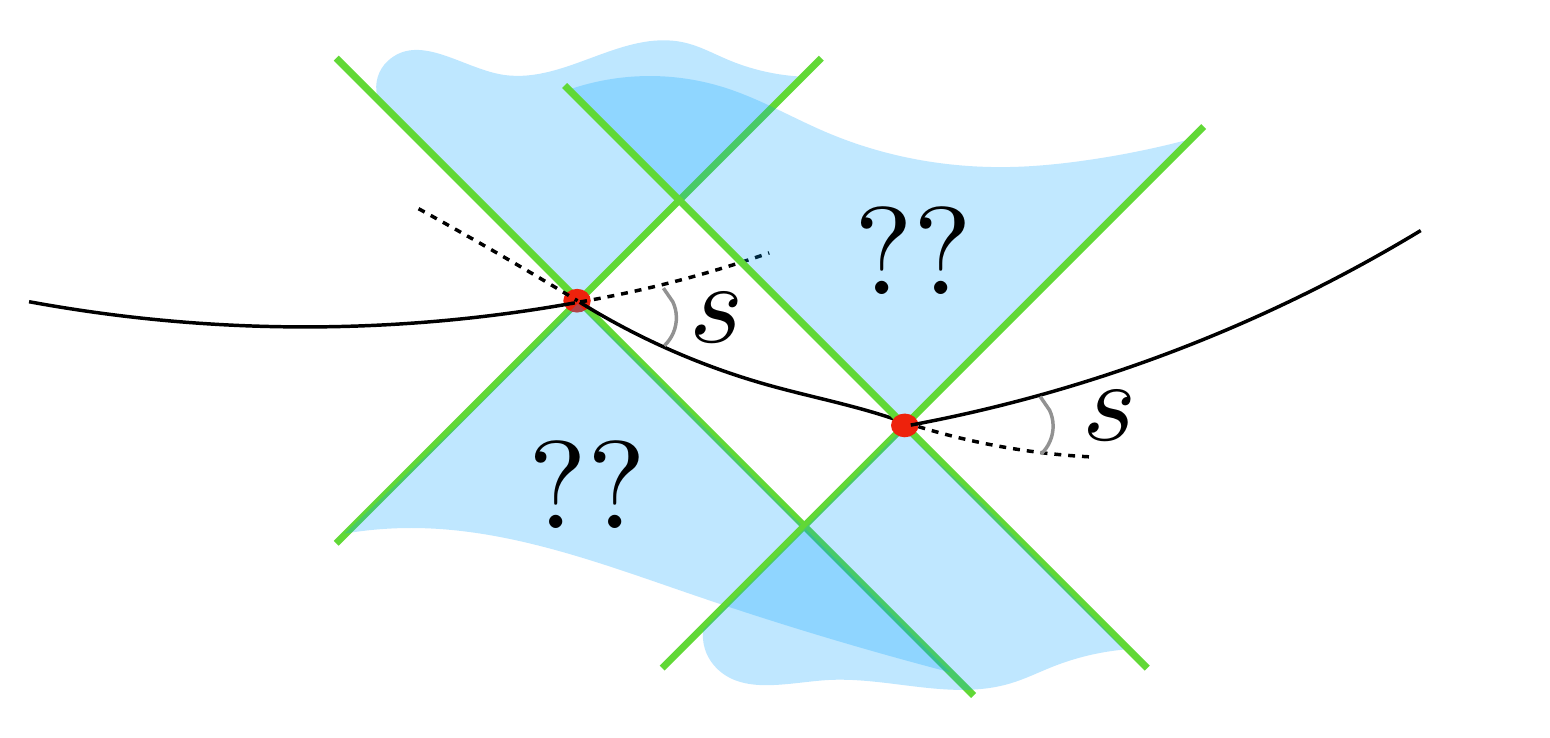}
\caption{\label{fig:double-two-sided} Double modular flow can also be computed if the parameters $x,y,s$ are tuned accordingly so that the geodesic threads itself through the various entanglement wedges. }
\end{figure}

In this section we consider modular flow for two nested regions. We will eventually study the one-sided correlator that was mentioned in the introduction, however let us start by considering the two sided correlator:
\be
\label{doubleside}
\left< \mathcal{O}(x) \Delta_B^{is} \Delta_A^{-is} \mathcal{O}(y) \right>
\ee
where $x \in \mathcal{D}(B)$ and $y \in \mathcal{D}(\bar{A})$. We expect the rules are the same for this correlator as for the single modular flow. Thus if we can arrange for a configuration of boundary points $x,y$ such that the picture of kinked geodesics in Figure~\ref{fig:double-two-sided} satisfies the boost condition in (\ref{match1}-\ref{match2}) at both $m_A$ and $m_B$,
then we can compute the correlator of \eqref{doubleside} in terms of a sum of the three piecewise lengths of the geodesic. Now we would like to freely change $s$ holding fixed $x,y$. As with the single modular flow case, this is generally not possible and we would need to move the boundary points as we change $s$ at the same time. We could again try to use the modular curves $x_{(y',\xi_B)}(s)$ 
and $y_{(x',\xi_A)}(s)$, but instead of this we would choosing to examine the mirror points.

\begin{figure}[h!]
\centering 
\includegraphics[scale=.65]{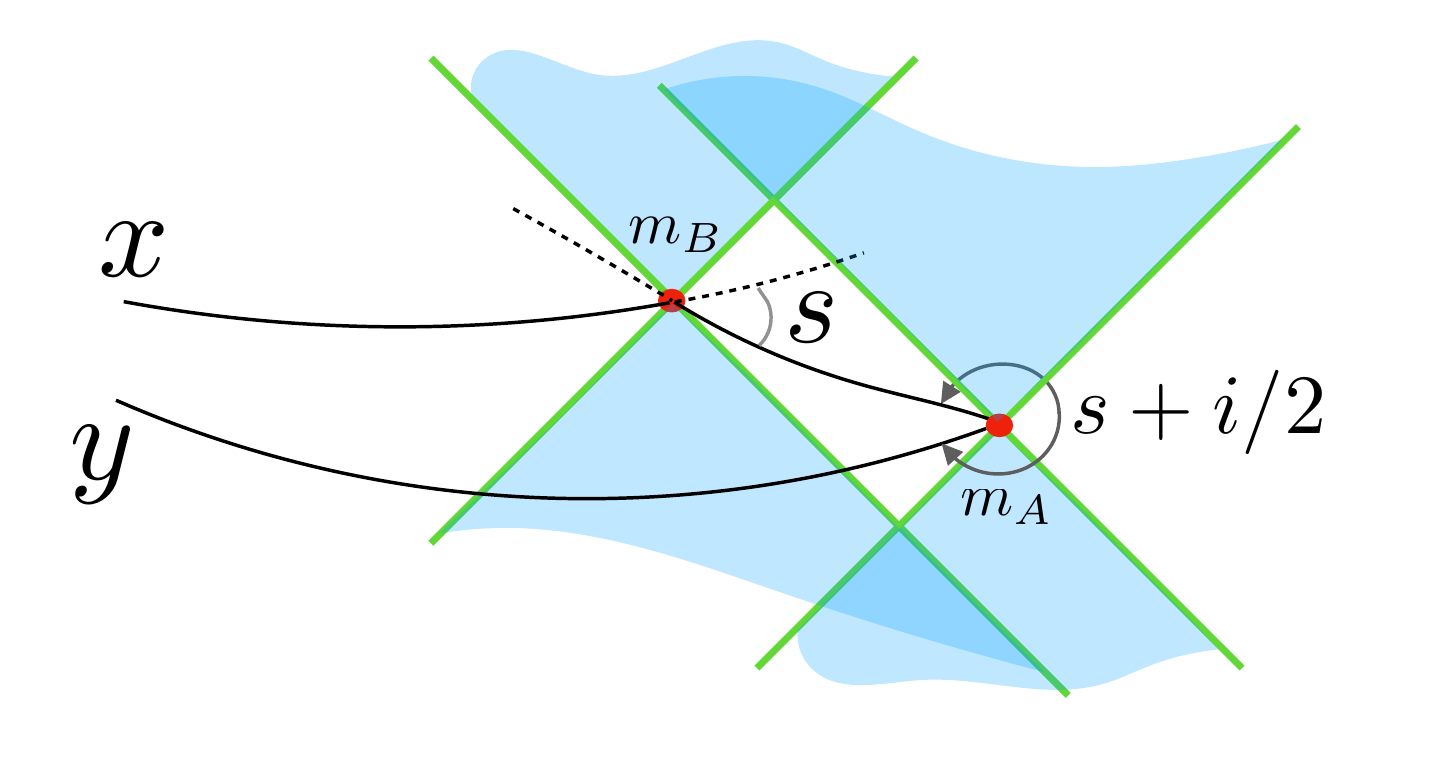}
\caption{\label{fig:double} Deforming the $A$ modular flow to include a Euclidean rotation by $\pi$ moves the geodesic back to the original side. We claim that the geodesic connected from $m_A$ to $y$ does not get effected by the non-trivial geometry of the Milne wedge associated to $m_B$. This can be understood by continuously deforming from the case in Fig~\ref{fig:double-two-sided}. } 
\end{figure}

To try to narrow the problem we move the points $x,y$ to the same boundary region $\mathcal{D}(A)$ and try to set $x=y$. As we do this we have to change the correlator to allow for an imaginary modular flow for one of the factors:
\be
\label{doublesideim}
\left< \mathcal{O}(x) \Delta_B^{is} \Delta_A^{-is+\theta} \mathcal{O}(y) \right>
\ee
We can imagine changing $\theta$ from $0$ to $1/2$ smoothly and as we do this we simultaneously change $y \rightarrow y_\theta$ to satisfy the conditions of Rule 1. We get a picture like Fig~\ref{fig:double}
where we note that, perhaps contrary to expectations, the reflected geodesic that comes from $m_B$ should not be effected again by the $\Delta_A$ modular operator and so it does not feel the non-trivial geometry in the Milne wedge associated to $m_A$ twice. Roughly speaking we should think of the reflected geodesic as living in a different/folded spacetime that does not contain the $s$-deformed Milne wedge time like from $m_A$. We will see this folded spacetime more explicitly when study these ideas in the replica trick.

\begin{figure}[h!]
\centering 
\includegraphics[scale=.7]{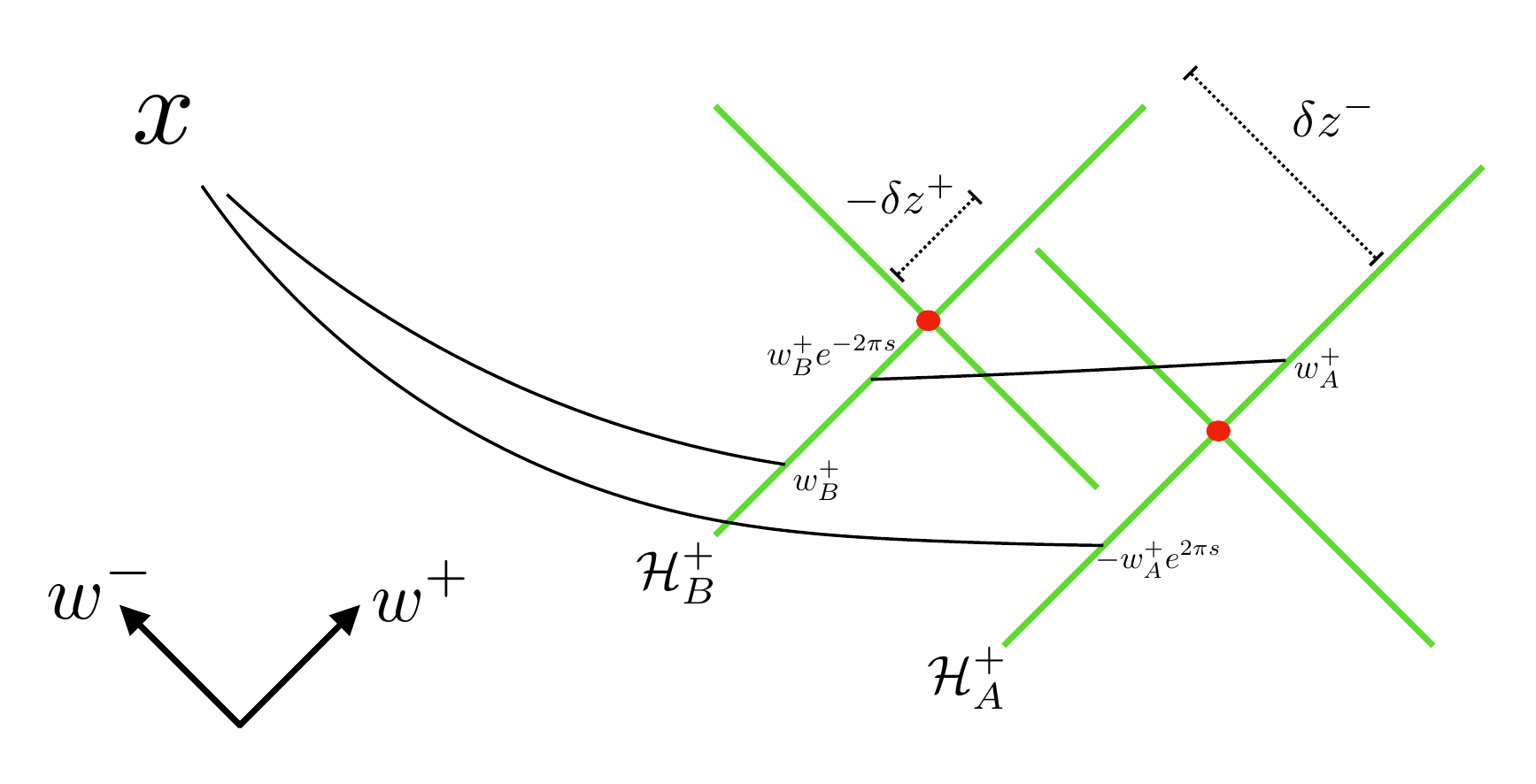}
\caption{\label{fig:calc} Computation of the double modular flow correlator for small deformations
$\delta z^\pm$. There are three segments and we extremize over parameters $w_A^+, w_B^+$ that have their origin at the respective entangling surfaces. }
\end{figure}

It is still however hard to have $x=y$ and satisfy all of the local boost rules for different values of $s$.  If we consider $B$ to be a small deformation of $A$, parameterized by $\delta z$ then we claim that we can approximately satisfy the local boost conditions up to $\mathcal{O}(\delta z)$ corrections. Thus we can apply rule $2$. The reason for this is as follows. When $\delta z=0$ and $x=y$, the correlator becomes exactly the mirror correlator studied in the previous section and any $s$ dependence goes away in this limit. This mirror correlator is determined by a geodesic from $x$ to $\xi(x)$. Deforming to non-zero $\delta z$ the local boost conditions slightly deform the location of the geodesics but overall the nested boosts will approximately reflect the geodesic back on itself. Applying the discontinuous version of the second rule \eqref{rule2disc} we arrive at a picture of geodesics that looks like Fig~\ref{fig:calc}. The picture is zoomed in close to the entangling surface at some point $\xi_\star$. The deformations of the entangling surface are defined in the normal coordinates of \eqref{gn}
where we will set $\delta z^\pm \equiv \delta z^\pm(\xi_\star)$. We pick the convention that $\delta z^\pm$ points from the $m_A$ to $m_B$, such that if nesting is satisfied $\delta z^+ < 0$ and $\delta z^- >0$.

The length of these three pieces of the curve can be computed to first order by expanding about the reflected geodesic associated to the $m_B$ entangling surface; from the boundary point $x$
and the reflection point $\xi_\star = \xi_B(x)$ . According to (\ref{rule2disc}) (\ref{vary}), the total length is:
\begin{align}\label{3.2.1}
2 \ell(x,X_B(\xi_\star)) & +  (n_+)_\star \left( w_B^+ - w_A^+ e^{2\pi s} -\delta z^+ \right)
+(n_- )_\star \delta z^- \non \\ &  + \left(\delta z^- (  -w_B^+ e^{-2\pi s} + w_A^+  -\delta z^+ ) + h_{ij} ( \delta \xi^i_A - \delta \xi^i_B ) ( \delta \xi^j_A - \delta \xi^j_B ) \right)^{1/2}
+ \mathcal{O}(\delta^2)
\end{align}
where we have used that fact that for the undeformed reflected geodesic $(\partial_\tau \xi^i )_\star =0$ and defined $\delta \xi_A = \xi_A - \xi_\star$ and $\delta \xi_B = \xi_B - \xi_\star$. Recall that 
$n$ was defined in \eqref{defn} and we only need evaluate it on the incoming side of the mirror geodesic. {It satisfies $(n_+ n_-)_\star = -1/4$.} The small parameter $\delta$ means any of: $\delta z^\pm, \delta \xi_A,\delta\xi_B, w_B^+, w_A^+$. Demanding (approximate) extermality with respect to $\xi_B,\xi_A, w^+_A$ and $w^+_B$ and by assuming that the quadratic terms are generically non-vanishing then we arrive at the relation:
\begin{align}
\delta \xi^i_A &= \delta \xi^i_B + \mathcal{O}(\delta^2) \\
2 e^{2\pi s} ( n_+ )_\star &= \frac{\delta z^-}{ \left( \delta z^- ( - w_B^+ e^{-2\pi s} + w_A^+ - \delta z^+ )  \right)^{1/2}}  + \mathcal{O}(\delta)\non\\
\end{align}
Note that this does not individually fix $w_A^+$ or $\delta \xi^i_A$ but it is not hard to see that they are systematically fixed by the higher order terms. The length of the combined curves is also fixed to the desired order:
\begin{align}
\label{terms}
&&= 2 \ell(x,X_B(\xi_\star)) + \frac{1}{2}  \left( -e^{-2\pi s_\star} (1+ e^{2\pi s}) \delta z^+ + e^{2\pi s_\star}   (1 - e^{-2\pi s}) \delta z^- \right)  \non\\
\end{align}
where
\be\label{numerator}
e^{2\pi s_\star} = \left(-\frac{ n_-}{ n_+} \right)_\star^{1/2}
\ee
This later quantity is related to the angle at which the probe geodesic comes into the entangling surface.

Note that there is a second way to arrive at \eqref{terms} that follows more directly the original version of Rule $2$. This is a little harder to describe, so we will not go into details. The idea involves allowing for small deformations of the entangling surfaces so that the local boost rules may be satisfied as we vary $s$. Expanding linearly about this deformation to come back to the original entangling surface gives the same result in \eqref{terms}.

With this result we can study the particular ratio of correlators that we mentioned in the introduction.
We will present theses slightly differently to make some of the properties obvious. Define the operator:
\be
U(t) = \Delta_B^{it} \Delta_A^{-it}
\ee
The operator norm of $U(t)$ is bounded by $1$ in the strip $ 0 < {\rm Im} t < 1/2$. 
For a recent review that discusses this property of nested modular operators see \cite{Witten:2018zxz}, some of the original literature can be found in \cite{zbMATH00845667,buchholz1990nuclear,araki2005extension}. We now calculate matrix elements of $U(t)$ using the normalized states:
\be
\left| \beta \right> = \frac{\Delta_B^{1/4} \mathcal{O}(x) \left| \psi \right>}{ \left< \psi \right|
\mathcal{O}(x) \Delta_B^{1/2}\mathcal{O}(x) \left| \psi \right> }  \qquad \left| \alpha \right> = \frac{\Delta_A^{1/4} \mathcal{O}(x) \left| \psi \right>}{ \left< \psi \right|
\mathcal{O}(x) \Delta_A^{1/2}\mathcal{O}(x) \left| \psi \right> }
\ee
So that we know that the function:
\be
i \mathcal{M}(t) +1\equiv \left< \beta \right| U(t) \left| \alpha \right>
\ee
is bounded by $1$ such that ${\rm Im} \mathcal{M}(E) \geq 0$ for $E\equiv e^{-2\pi t}$ in the lower half plane. This function has other known symmetries, including an interesting generalization of the KMS condition.  They are:
\begin{align}
\label{reality}
\mathcal{M}(t) = - \left(\mathcal{M}(t^\star + i/2)\right)^\star
\\
\label{swap}
\left. \mathcal{M}(t)\right|_{A \leftrightarrow B} = \mathcal{M}(t + i/2)
\end{align}
and follow from the usual manipulations that derive the KMS condition for a single modular flow, with the additional result that: 
\be\label{moduoperator}
J_B J_A = \Delta_B^{1/2} \Delta_A^{-1/2}
\ee
which follows since we have the identity:
\be
J_B \Delta_{B}^{1/2} J_A \Delta_A^{1/2} \mathcal{O}_B \left| \psi \right> =  \mathcal{O}_B \left| \psi \right> 
\ee

We will now calculate this for small deformations between $A$ and $B$. $\mathcal{M}$ is related to the double flow correlator we computed above if we set $s = t - i/4$. Since we defined everything with respect to the mirrored operator associated to $B$, the $A$ mirror operator that appears in the denominator of $\mathcal{M}$ receives corrections that cancel the terms without any $s$ dependence in \eqref{terms}.
We are left with:
\be
\label{finalM}
\mathcal{M}(t) = - \frac{m}{2} \left(- \delta z^-(\xi_\star) e^{-2\pi(t-s_\star)} + \delta z^+(\xi_\star) e^{2\pi(t-s_\star)} \right) + \mathcal{O}(\delta z^2)
\ee

Note that this satisfies the reality condition of \eqref{reality} and also the condition \eqref{swap} assuming we can achieve the switch between regions to linear order ing $\delta z^\pm$
by simply switching the sign $\delta z^\pm \rightarrow - \delta z^\pm$.

It is now clear that this violates the boundedness of $U(t)$ unless:
\be
\label{finewn}
\delta z^-(\xi_\star) > 0 \qquad \delta z^+(\xi_\star) < 0
\ee
Recall that the point $\xi_\star$ is determined by the mirrored geodesic at the $B$ entangling surface $\xi_B(x)$. So as we vary $x$ we can map out the relative entangling surface location, and in particular this location is constrained by Entanglement Wedge Nesting \eqref{finewn}.
Note that if we further maximize the correlator over real $s$ with $t = s + i/2$, at the maximum point $s = s_{max}$ this correlator simply computes the length of bulk deformation:
\be
\mathcal{M}\mathcal(s_{max} - i /2) = m \sqrt{ - \delta z^-(\xi_\star) \delta z^+(\xi_\star) }
\ee
We thus expect that a fairly detailed picture of the bulk can be obtained from these correlators.
Note that EWN can be shown to follow directly from the bulk theory after assuming the Null Energy Condition (NEC) and a classical description via Einstein's equations \cite{Wall:2012uf}. Here we have given a boundary argument, a goal that we have not yet realized, would be to find a boundary description of the the bulk NEC. However the relation between EWN and NEC is not direct so we cannot use our methods in their current form. We still consider this a step in the right direction.

We also note that EWN in some sense is a trivial consequence of the JLMS map between bulk and boundary modular Hamiltonians. We know the double modular flow correlators of the boundary must have a certain analytic structure dictated by the causality of nested regions, and since one can be reasonably sure that any bulk violation of nesting would lead to non-analyticities of the bulk modular flow correlators, violations of the EWN would be in conflict with the JLMS map. Despite this our specific results are useful since they tell us exactly how causality and analyticity would be violated. Not only that we can use these results to extract information about the bulk directly from the boundary theory. 

\subsection{The Quantum Null Energy Condition}

It is well known that the EWN condition, when applied to nested regions near the boundary of $AdS$ reduces to the statement of the Quantum Null Energy Condition \cite{Bousso:2015mna,Bousso:2015wca} for the dual boundary QFT \cite{Koeller:2015qmn}. Thus it must be the case that our new purely boundary understanding of EWN reduces to the recent discussion of the QNEC purely from the boundary interacting QFT point of view \cite{Balakrishnan:2017bjg}. The ingredients are basically the same, so there should be no surprises here. It is still useful to go through the exercise, as it will give an independent check on the rules that we describe.
Here we will compute $\mathcal{M}$ in a similar kinematical limit to \cite{Balakrishnan:2017bjg}. For a more precise match to the function $f(s)$ in that paper see Appendix~\ref{app:qnec}.

To extract the QNEC the bulk point $\xi$ must limit to the boundary of the entangling surface. This will happen if $x \rightarrow \partial B$ on the boundary. In this case we are probing physics near the UV of the bulk theory and so we can take the bulk space to be $AdS$ plus some small corrections that die off near the boundary in a Fefferman-Graham expansion. For simplicity we will take the boundary theory to be a CFT (with no mass deformations) and assume the entangling surface near the point $x$
 is sufficiently flat that we can ignore extrinsic curvature effects. 
 
 We aim to compute \eqref{finalM}. There are two ingredients to this, relating to the motion of the entangling surface, and the probe geodesic. We start by finding $\delta z^\pm$. In some general bulk state $\psi$, the metric near the boundary has a Fefferman-Graham expansion,
\begin{eqnarray}\label{3.3.1}
ds^2=\frac{1}{z^2}\bigg(-dudv+dz^2+d\vec{y}^2+\frac{16\pi G_N}{d}z^d\la T_{\mu\nu}\ra_{\psi}dx^{\mu}dx^{\nu}\bigg)
\end{eqnarray}
We choose the RT surfaces, $m_A$ and $m_B$, to be anchored at $\partial A$ and $\partial B$ which are taken to be lightlike separated from each each along $v=0$ and nearly light like separated from $x$. We place the entangling surfaces symmetrically about the origin $u=0$ 
\begin{eqnarray}\label{3.3.3}
u_{\partial A}=(X^-_0)_A=-\frac{\delta x^-}{2},\qquad u_{\partial B}=(X^-_0)_B=\frac{\delta x^-}{2}
\end{eqnarray}  
In the bulk the co-dimensional two RT surfaces can be parameterized by $u,v=X^{\pm}(z,\vec{y})$ where $X^{\pm}_0=X^{\pm}(0,\vec{y})$ with an expansion near the boundary:

\begin{eqnarray}\label{3.3.2}
X^-_{A,B}=(X^-_0)_{A,B}+\OO(z^2),\qquad X^{+}_{A,B}=\frac{8G_N}{d}z^d \mathcal{P}_-^{A,B}(y)
\end{eqnarray}
The RT surfaces are lightlike separated on the boundary, and become spacelike separated in the bulk.

The deformation vector between the two entangling surface is:
\begin{eqnarray}
\delta z=\delta x^- \partial_{u}+\frac{8 G_N}{d}z^d(\mathcal{P}_-^B-\mathcal{P}_-^A) \partial_{v}
\end{eqnarray}
and in order to find $\delta z^\pm$, which were defined naturally in Gaussian normal coordinates we need to decompose this vector into the null normal vectors $k_{\pm}$ (and possibly some  vector along the surface). The null normals are linear combination of normal vectors defined via $n^a=g^{-1}d f^a$ for $a=\pm$ and:
\begin{eqnarray}\label{1.2}
f^-=u-X^-(z,\vec{y}),\qquad f^+=v-X^+(z,\vec{y})
\end{eqnarray}
which describe the location of the entangling surface as the solution to $f^\pm =0$. We find two unnormalized normal vectors,
\begin{eqnarray}
n_1=-z^2 \partial_{v},\qquad n_2=-z^2 \partial_{u}-2G_N z^{d+1}\mathcal{P}_-^A\partial_z-\frac{16\pi G_N}{d}\la T_{uu}\ra_{\psi} z^{d+2}\partial_v
\end{eqnarray}
where the null vectors $k_{\pm}=(h_a n^a)_{\pm}$ are linear combination of $n^a$ which satisfies $(k_{\pm})^2=0,~k_+\cdot k_-=-1/2,~k_{\pm}\cdot {\Pi_i}=0$, accordingly, one can solve for $k_{\pm}$

\begin{eqnarray}\label{2.4}
k_+=z \partial_{v}+ \ldots\,,\quad k_-=z\partial_{u}+\frac{16\pi G_N}{d}\la T_{uu}\ra_{\psi}z^{d+1} \partial_{v}+ \ldots
\end{eqnarray}
Thus we can find $\delta z^{\pm}$,
\begin{eqnarray}
\delta z^-=z^{-1}\delta x^-,\quad \delta z^+=\frac{16\pi G_N}{d}z^{d-1}\bigg(\frac{\mathcal{P}^B_+-\mathcal{P}^A_+}{2\pi}-\delta x^- \la T_{uu}\ra_{\psi}\bigg)
\end{eqnarray}
We now have to treat the dependence on the probe geodesics which is completely encoded  $\mathcal{M}$ via $e^{2\pi s_\star}$. In more general coordinates we have:
\begin{eqnarray}
e^{-4\pi s^*} = - \frac{\partial_{\tau}x(\tau)\cdot k_+}{\partial_{\tau}x (\tau)\cdot k_-}\Lvert_{\tau_m} \end{eqnarray}
Now since we want to keep track of the leading term multipying each of $e^{\pm 2\pi s}$ in the $G_N$ expansion of $\mathcal{M}$, we can simply estimate $e^{-4\pi s_\star}$ in vacuum AdS. In this case since  the geodesic must reach the entangling surface $m_B$ perpendicularly the geodesic must be a half of the semi circle centered on the entangling surface.  In vacuum AdS, we know that the projection of the geodesic in $u-v$ plane is a straight lines, thus:
\begin{eqnarray}\label{2.5}
e^{ - 4 \pi s^\star} =-\frac{\Delta u}{\Delta v}+\OO(G_N)
\end{eqnarray}
We can also estimate the point where geodesic intersect $m_{B}$ in the bulk using vacuum AdS results, $z_{\rm max} =\sqrt{-\Delta u\Delta v/4}$.  We can obtain $\mathcal{M}$ according to (\ref{finalM}),
\begin{eqnarray}
\label{finres}
 i\mathcal{M} \simeq e^{-2\pi s}\frac{m}{\Delta u}\delta x^--e^{2\pi s }\frac{16\pi G_N m}{d\Delta v}\left(-\frac{\Delta u\Delta v}{4}\right)^{d/2}\bigg(\frac{\mathcal{P}^B_--\mathcal{P}^A_-}{2\pi}-\delta x^- \la T_{uu}\ra_{\psi}\bigg)
\end{eqnarray}

Positivity reduces to the statement of the QNEC. Note that the first term is trivially positive and actually swamps the second term so this is a bit subtle here. The fix for this is to write a sum rule for these terms, that projects onto each individual term, following \cite{Hartman:2015lfa,Hartman:2016lgu,Hofman:2016awc}. However we will not go into these details.
Note that, although we will not show the calculations, the same result \eqref{finres} can be found using the boundary theory defect CFT  like calculation of \cite{Balakrishnan:2017bjg} or the algebraic computations in \cite{Ceyhan:2018zfg} adapted to the slightly different correlator studied here.

\section{Justification of rules}
\label{derivation}

\subsection{$\Delta^{1/2}$}
\label{dhalf}

We will start with an interesting observation for computing the specific modular flow corresponding to a $\pi$ rotation. That is, the two point function of an operator with its mirror.  This will then lead us to the more general case. 

In terms of reduced density matrices associated to the $A$ Hilbert space consider the following correlator:
\be
\label{even}
G_n = {\rm Tr}_A \rho^{n/2}_A \mathcal{O}(x) \rho^{n/2}_A \mathcal{O}(x) \Big/ {\rm Tr} \rho_A^n
\ee 
This can be computed using a path integral for \emph{even} integers $n=2m$. If we can then find a natural analytic continuation in $n$ the mirror correlator can be extracted as:
\be
\left< \mathcal{O}(x) \Delta^{1/2}_A \mathcal{O}(x) \right>
= {\rm lim_{n \rightarrow 1}} G_n 
\ee
Such continuations from even integer to $n=1$ are not unprecedented and have come up in CFT replica computations of entanglement negativity \cite{calabrese2012entanglement}. Here we will find a natural analytic continuation in the dual gravitational theory.

\begin{figure}[h!]
\centering 
\includegraphics[width=.4\textwidth]{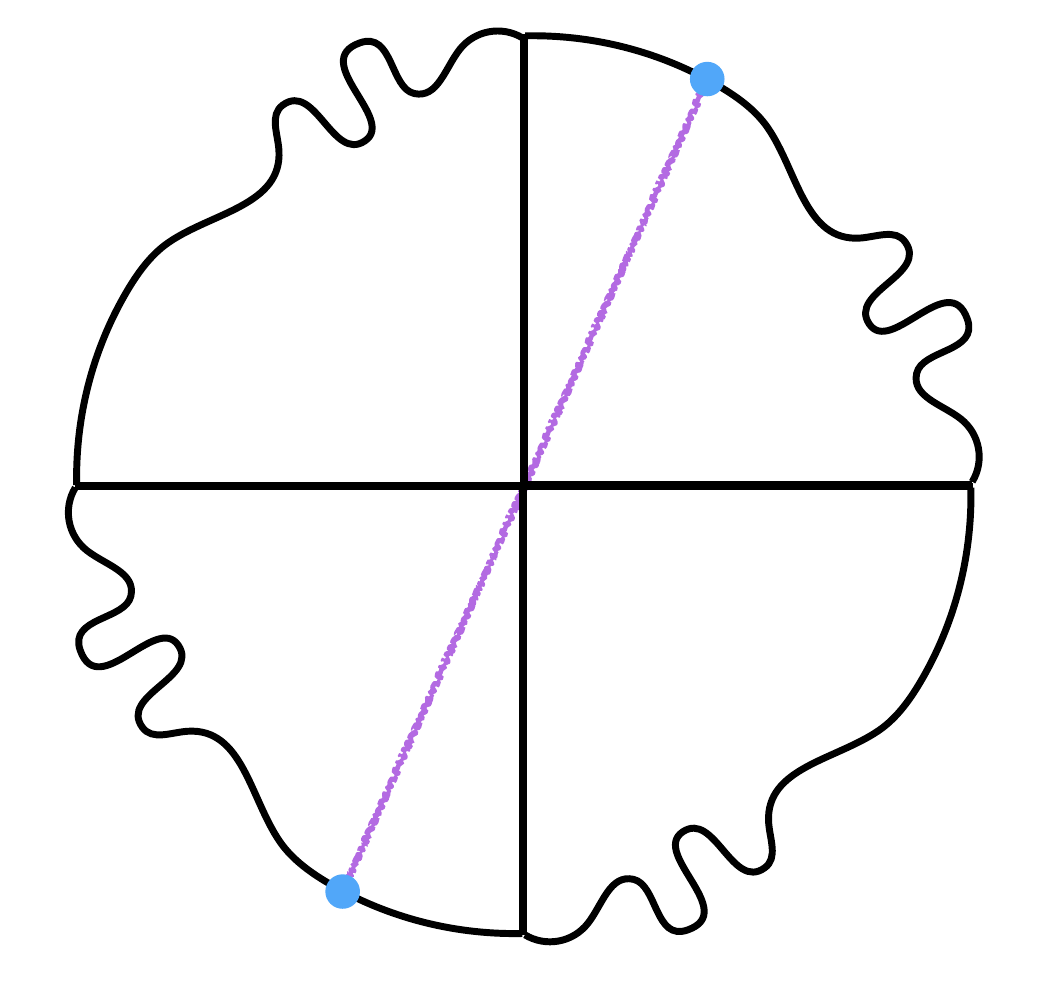}
\hspace{1.5cm}
\includegraphics[width=.4\textwidth]{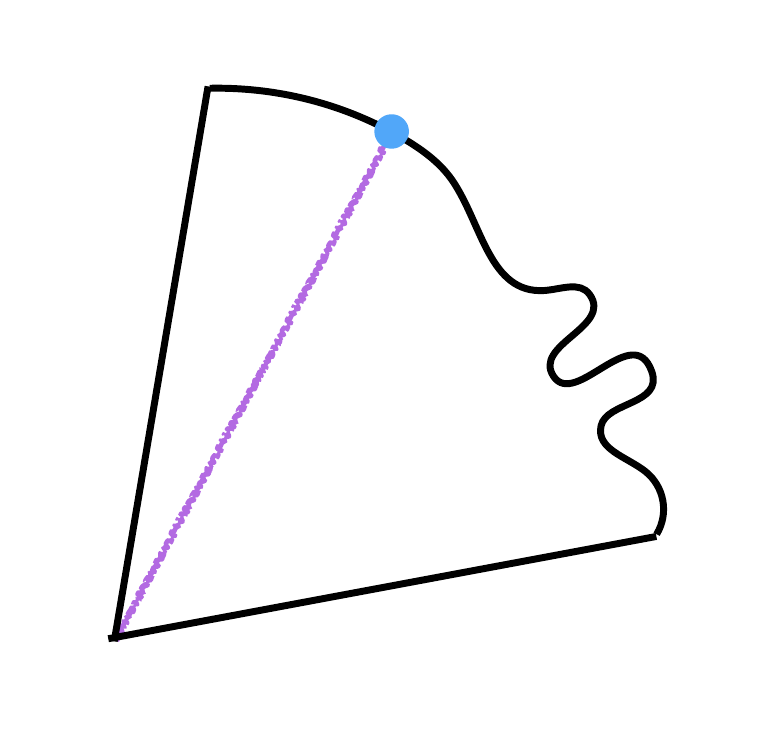}
\caption{\label{fig:LMeven} The strategy for computing correlators with $\Delta^{1/2}$ involves computing correlators in the smooth $n$-replicated geometry. Taking $n$ even one always finds a geodesic that passes through the $\mathbb{Z}_n$ fixed point. Analytically continuing this answer results in a geodesic ending on the conical singularity spacetime to the right. The correlator is related to twice the length of this geodesic. }
\end{figure}

The replica trick for entanglement computations in holography was developed in \cite{Faulkner:2013yia,Lewkowycz:2013nqa}. The Lewkowyc-Maldacena (LM) picture of the bulk spacetime, without any operators insertions, is one where the classical bulk spacetime $\mathcal{B}_n$ is smooth around the fixed point of the $\mathbb{Z}_n$ symmetry which we assume extends into the bulk from $\partial A$. Recall that $\mathcal{B}_n$ solves Einstein's equations and has boundary $\mathcal{M}_n$ the replicated boundary theory.  The locus of $\mathbb{Z}_n$ fixed points is co-dimension two and upon analytic continuation as $n\rightarrow 1$ becomes the entangling surface $m_A$. Focusing on a slice through this fixed point locus a natural cartoon of the bulk is shown in Figure~\ref{fig:LMeven}

In order to compute the correlator in \eqref{even} we should simply solve the Klein-Gordan wave equation in this spacetime with boundary sources at the operator locations. The large $m$ approximation computes the answer in terms of the length of a geodesic that passes between the two boundary points. Since $n=\,$even, the two operator insertions are necessarily on opposites sides of the above picture. They are related to each other by a $\mathbb{Z}_2$ symmetry which is the subgroup generated by $\tau^{n/2} \in \mathbb{Z}_n$ for $\tau$ the cyclic permutation. Assuming the bulk geodesic also satisfies this $\mathbb{Z}_2$ symmetry (this assumption is similar to the assumption of the lack of replica symmetry breaking in the original work of \cite{Faulkner:2013yia,Lewkowycz:2013nqa}) then it necessarily passes through the fixed point locus of the $\mathbb{Z}_2$ symmetry which is the entangling surface. 

The correlator is approximately:
\be
\label{agn}
g_n = {\rm min}_\xi \exp\left( - 2 m \ell_n(x,X^n_A(\xi)) + \ldots \right)
\ee
where $\ell_n(x,X^n_A(\xi))$ is the length of a geodesic in bulk $\mathcal{B}_n$ from the boundary point $x$ on the first replica to the $\mathbb{Z}_n$ fixed point locus denoted here by $X_A^n(\xi)$. Minimization guarantees that the geodesic is smooth when we join it with the rotated geodesic after an action of $\tau^{n/2}$ and the symmetry means that it has the same length, hence the factor of $2$ above. The minimizing point will depend on $n$ and will be denoted $\xi_n(x)$. 

It is then not hard to guess the analytic continuation of \eqref{agn}. Following LM we introduce a new spacetime $\widehat{\mathcal{B}}_n = \mathcal{B}_n/\mathbb{Z}_n$ which has a conical deficit at $X_A^n$ but allows for a definition of the bulk when $n$ is not integer. LM argued that $n$ times the gravitational action of $\widehat{\mathcal{B}}_n$, removing any contribution from the conical singularity, computes the Renyi entropies:
\be
{\rm Tr} \rho_A^n = \exp\left( - n G_N^{-1} S_{\rm grav}( \widehat{\mathcal{B}}_n ) + \ldots \right)
\ee
which is well defined for non-integer $n$ and agrees with the action of $\mathcal{B}_n$ for integer $n$.

In the spacetime $\widehat{\mathcal{B}}_n$ we can still find a geodesic from $x$ to the conical singularity $X_A^n(\xi)$ and minimize it's length. Denoting the length of this geodesic also by $\ell_n$ but now allowing for $n$ non-even integer gives the natural continuation of \eqref{agn}. The limit $n \rightarrow 1$ gives \eqref{mirrorcorr} where it should be clear that $\lim_{n \rightarrow 1} \xi_n(x) = \xi(x)$.

Note that just like the LM trick where the new spacetime $\widehat{\mathcal{B}}_n$ only solves the equations of motion away from the conical deficit, the curve constructed above also only solves the geodesic equations of motion away from the conical singularity. Unfortunately this method cannot be used to compute the backreaction of a very heavy bulk geodesic, if say $m \approx 1/\ell_p$ since the back reaction would destroy the underlying $\mathbb{Z}_n$ symmetry that we are using to do the continuation. A similar issue would arise if one tried to compute the one point function of heavy (single replica) operators using the LM spacetime, since they would have a large back-reaction on the bulk that does not preserve the cyclic symmetry. 

Since this result is rather robust it would be nice to understand this from different points of view. Indeed the paper \cite{aitornew} that has similar results to ours discusses a completely different method to justify their results using the modular zero modes discussed in \cite{Faulkner:2017vdd}. They did not apply this to the $\Delta^{1/2}$ case, but this does seem possible and it would be worth developing further the relation between our two complementary approaches. 

\subsection{Modular flow replica trick}

We would like to generalize the above to compute the following:
\be
\label{gnk}
g_{n,k}(x,y) = {\rm Tr} \rho^{n-k}_A \mathcal{O}(x) \rho^k_A \mathcal{O}(y)
\ee
which can be computed with a path integral for $n,k$ integers and $0 \leq k \leq n$. 
We will work with $x,y$ inserted in Euclidean on one of the boundary theory replicas. 
In Euclidean it is easiest to understand operators inserted on $\in A, \bar{A}$ since they can be easily continued from real time, \footnote{The arguments basically assume we have a natural Euclidean extension of the semi-classical bulk spacetime dual to the state $\psi$, however our results are so general that this is likely not necessary. } but more generally one should define:
\be
\rho \mathcal{O}(x)
\ee
as the path integral that produces $\rho$ except with an additional operator insertion at $x$.
The final goal is to compute \eqref{gnk} and then find a natural analytic continuation for $n$ real and $k$ complex  such that we can eventually send  $n \rightarrow 1$ and $k \rightarrow  is$. This continuation needs to proceed in a particular order, spelled out in the Appendix~D of \cite{Balakrishnan:2017bjg}.

This correlator obeys a KMS condition even after continuation:
\be
\qquad g_{n,k+n}(x,y) = g_{n,-k}(y,x)
\ee

 We now proceed to map this correlator to the bulk. Again in the spacetime $\mathcal{B}_n$ we solve for a boundary to boundary propagator by solving the Klein-Gordan equation with fixed boundary sources at $x^{(0)}$ and $y^{(k)}$ where the superscript denotes which replica to source the boundary operator. This is the boundary limit of the bulk to bulk propogator - which is simply the Green's function for the Klein-Gordan equation where there are no ordering ambiguities because we work in Euclidean. This Green's function will play an important role below and we denote it $G_n(X^{(k)},Y^{(q)})$ where $X,Y$ are coordinates on a single bulk replica and $k,q$ label the different replicas. Then schematically:
\be
g_{n,k}(x,y) = \lim_{X \rightarrow x, Y \rightarrow y} G_n(X^{(0)},Y^{(k)})
\ee
where the details of this limit, and stripping of the factors of the AdS radial direction $z$, are not important. 

Firstly we note that there is already a natural analytic continuation of $G_n$ which we describe now. Since we are essentially studying free bulk quantum field theory we can slice up the space $\mathcal{B}_n$ using angular like coordinates around the entangling surface $m_A$ (the $\mathbb{Z}_n$ fixed points) and re-interpret this in a Hilbert space langauge for the bulk sub-region $a$ which is a region between $A$ and $m_A$. That is:
\be
\label{gnreal}
G_n(X^{(0)}, X^{(k)}) = {\rm Tr}_a \rho_{(n)}^{n-k} \phi(X) \rho_{(n)}^{k} \phi(Y)
\ee
where $\rho_{(n)}$ is defined via a path integral with boundary conditions for the quantum fields above and below a single replica: at $a^{(0)}$ and  $a^{(1)}$ where the superscript again denotes the particular replica where the sub-region $a$ lives.  This density matrix was introduced in \cite{Faulkner:2013ana} in order to compute the quantum corrections to the RT formula. The sub-script $(n)$ is to remind us that this path integral depends explicitly on $n$ via the bulk spacetime $\mathcal{B}_n$. This reduced density matrix can also be defined on the spacetime $\widehat{\mathcal{B}}_n$ where $n$ need not be an integer. Such that:
\be
g_{n, is} =  \lim_{X \rightarrow x, Y \rightarrow y} {\rm Tr}_a \rho_{(n)}^{n-is} \phi(X) \rho_{(n)}^{is} \phi(Y)
\ee
is the natural analytic continuation. Recall that because we are working away from $n=1$ modular flow is analytic and periodic in the strip $ 0 \leq {\rm Im} s \leq 2\pi n$.  This answer should come as no surprise. It basically reduces to  statement that bulk modular flow $=$ boundary modular flow when $n=1$.   

It is also not a very useful answer from our point of view since it relates a very hard problem to a hard problem. However we will need the various quantities defined above in later steps. 

Instead we will return to integer $n,k$ for a while. The next step is to apply Green's theorem on $\mathcal{B}_n$ to breakup the  Green's function $G_n$ into an integral over several Green's functions. This will allow us to isolate various contributions to $G_n$ in a way that allows us to do the analytic continuation.

\begin{figure}[h!]
\centering 
\includegraphics[width=.38\textwidth]{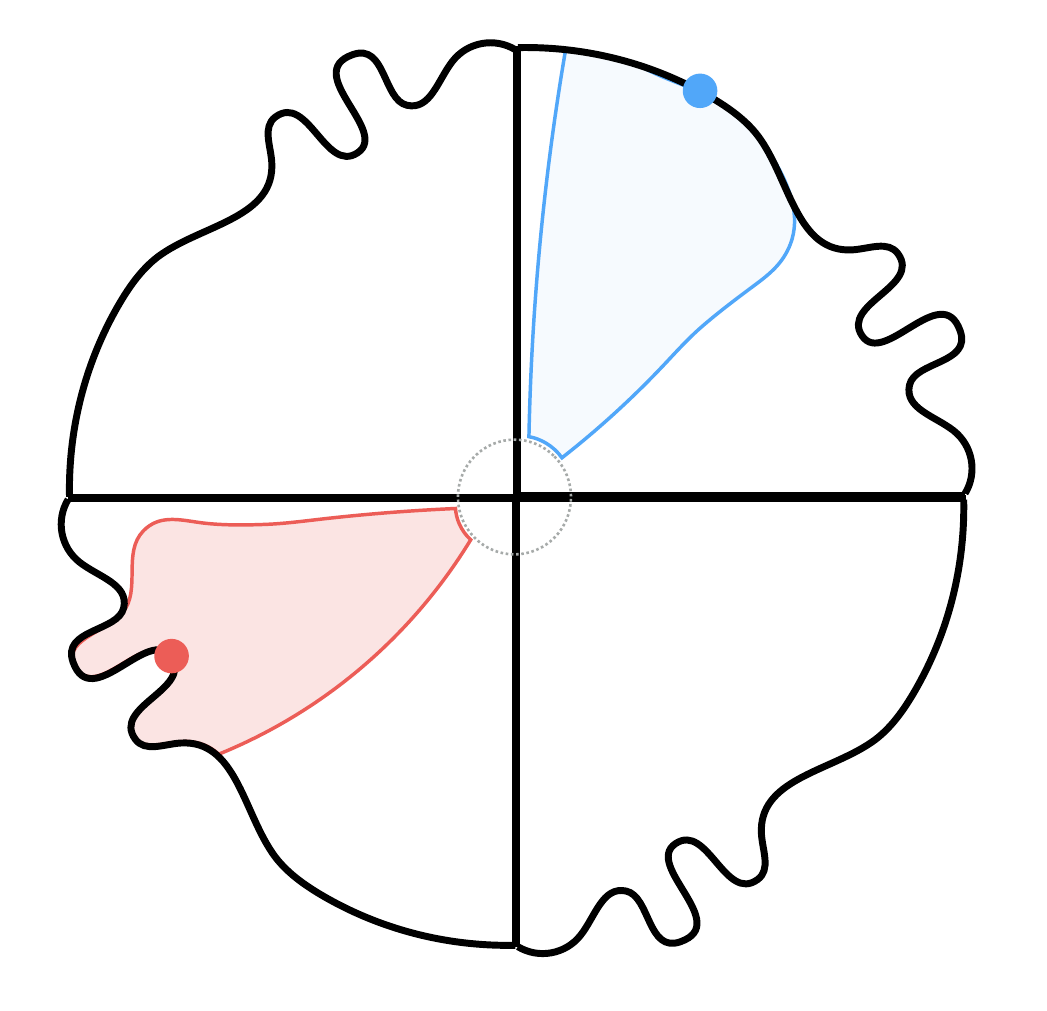}
\hspace{2cm}
\includegraphics[width=.38\textwidth]{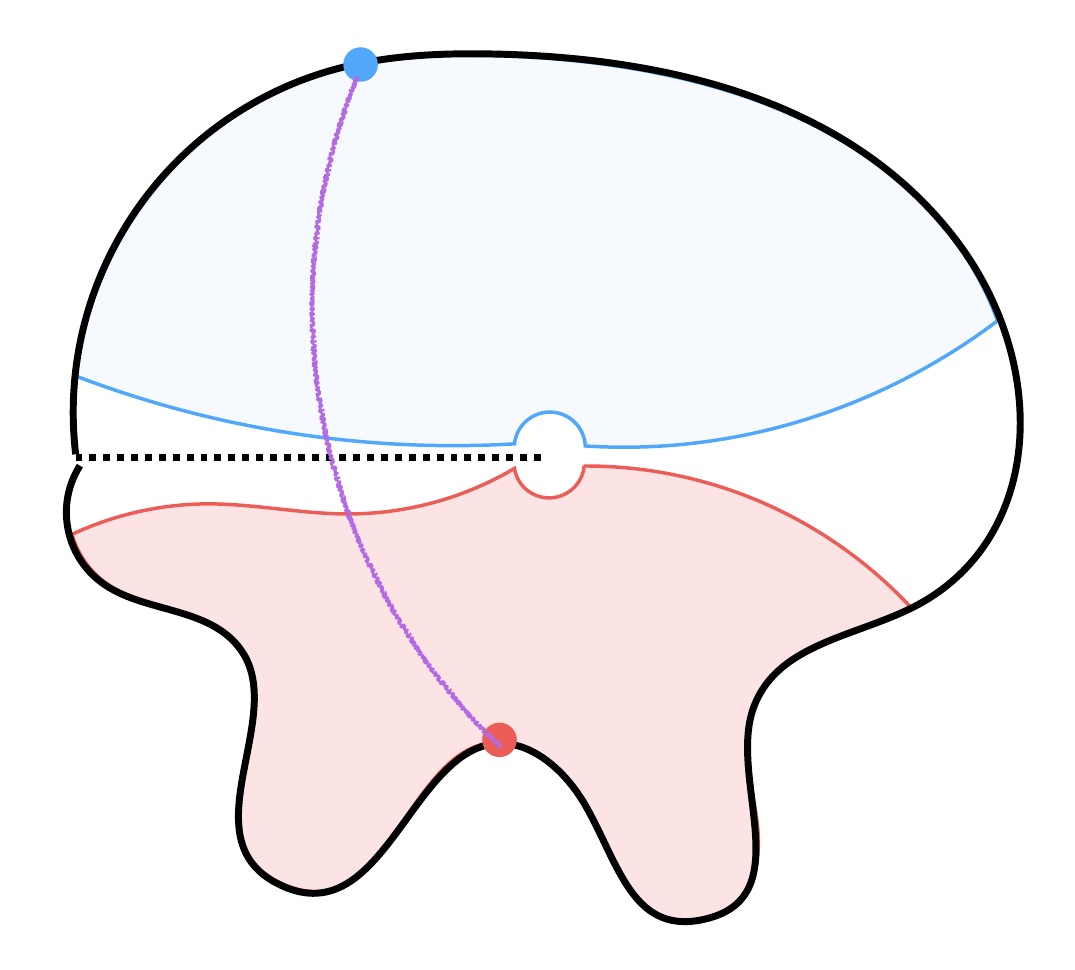}
\caption{\label{fig:LMregions}We can compute a two point function in the replicated geometry by splitting the contribution into three $S_1, S_2$ and $(S_1 \cup S_2)^c$. We pick $S_{1,2}$ to lie on a single replica separated by $k$ replicas. We pick the regions to come close to the $\mathbb{Z}_n$ fixed point, but drill a hole out of the region there. The left shows the case with $n=4$ and the right is $n=1$. We pick the two regions to be non-intersecting when viewed on a single replica. The regions will depend on $n$, and we can define these regions on the conical space for non-integer $n$. }
\end{figure}

Consider the following general identity on Green's function on some space $S$
\be
G(X,Y) = \int_{\partial S_1} G(X,Z_1) n_1\cdot \left.\mathop{\nabla}^\leftrightarrow\right._{Z_1} G(Z_1,Y)
\ee
where $S_1 \subset S$ is some region containing $X$ and not $Y$. Applying this rule again we have:
\be
G(X,Y) = \int_{\partial S_1} \int_{\partial S_2} G(X,Z_1) n_1\cdot \left.\mathop{\nabla}^\leftrightarrow\right._{Z_1} G(Z_1,Z_2) n_2\cdot \left.\mathop{\nabla}^\leftrightarrow\right._{Z_2} G(Z_2,Y)
\ee
where $X \in S_1$ and  $Y \in S_2$ and $S_1 \cap S_2 =0$. We will now apply this to the bulk space $\mathcal{B}_n$ where we pick two non interesecting regions $S_{1},S_{2}$ which contain the two boundary points $x,y$ on replica $0$ and $k$ respectively. We have:
\be
\label{comp}
g_{n,k}(x,y) = \int_{\partial S_1 } \int_{\partial S_2 }   K_n(x^{(0)},Z^{(0)}_1) n_1\cdot \left.\mathop{\nabla}^\leftrightarrow\right._{Z_1}G_n(Z_1^{(0)},Z_2^{(k)}) n_2\cdot \left.\mathop{\nabla}^\leftrightarrow\right._{Z_2} K_n(Z_2^{(k)},y^{(k)})
\ee
where $K(x,Z) = \lim_{X \rightarrow x} G_n(X,Z)$ is the bulk to boundary propogator. It is important that we pick the regions $S_{1,2}$ to be contained in a single replica but to come close to the $\mathbb{Z}_n$ fixed point. See Figure~\ref{fig:LMregions} for further explanation of what the regions $S_{1,2}$ should look like.

We use
the replica symmetry to rewrite $K_n(Z_2^{(k)},y^{(k)}) = K_n(Z_2^{(0)},y^{(0)})$. The two bulk to boundary propogators make no reference to $k$ and so the analytic continuation is simple:
\be
\label{outside}
K_n(x^{(0)},Z^{(0)}_1) = {\rm Tr} \rho_{(n)}^n \phi(x) \phi(Z_1)
\qquad K_n(y^{(k)},Z^{(k)}) = {\rm Tr} \rho_{(n)}^n \phi(y) \phi(Z_2)
\ee
where now the only non-trivial analytic continuation is from the central term in \eqref{comp} which connects the two different replicas. Of course this is  analogous to the realtime discussion of Section~\ref{somerules} where modular flow only came from the connections across the two regions $\mathcal{E}_a$ and $\mathcal{E}_{\bar{a}}$.  However now we can use the same analytic continuation as in \eqref{gnreal} where we simply replace the bulk correlator with modular flow with respect to $\rho_{(n)}$:
\be
\label{midterm}
G_n(Z_1^{(0)}, Z_2^{(k)}) = {\rm Tr}_a \rho_{(n)}^{n-is} \phi(Z_1) \rho_{(n)}^{is} \phi(Z_2)
\ee
Again one might not suspect this is useful since we  don't know much about bulk modular flow, however \emph{if we can show that both the integrals over $\partial S_{1,2}$ in \eqref{comp} are dominated near the $\mathbb{Z}_n$ fixed point} then we can approximate the modular flow correlator, for example when $s = i \theta$, by a simple geometric rotation around the fixed point. Of course for imaginary modular flows, in Rindler space for the bulk QFT in vacuum we know the answer is given by such a rotation. So here we are using the fact that the entanglement structure in QFT is universal at short distances where we can additionally neglect the shape of the entangling surface and simply take it to be flat. 

While we do not know of a general proof of this statement, it seems likely a general argument can be made along the lines of \cite{Hollands:2018wzi,fredenhagen1985modular}. For now we will be content to look at some example computations where corrections away from Rindler flow were explicilty calculated using defect CFT methods \cite{Balakrishnan:2017bjg}. There one finds that when both operators are near the entangling surface the geometric flow receives corrections in powers of the distance to the entangling surface. The powers are determined by the various operators that live on the defect CFT for the $n$-replicated theory. Most of these corrections do not come with powers of $m$ such that  they are negligent compared to the leading semiclassical piece and deformations thereof. Note that if we are using the CFT, we are examining the correlation function at scales much shorter than $m^{-1}$. Of course there will also be corrections involving the mass, however we expect these are trivial since we know that in Rindler space the massive theory also has geometric modular flow. 

\begin{figure}[h!]
\centering 
\includegraphics[width=.35\textwidth]{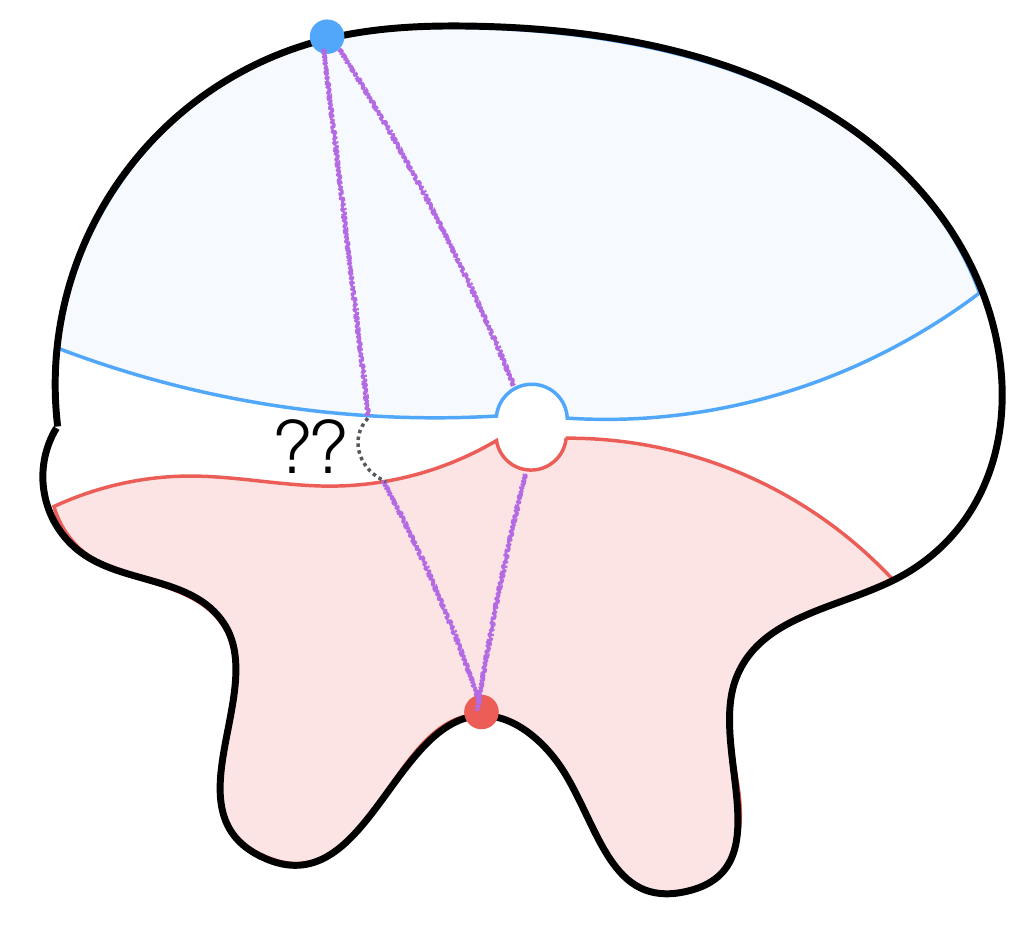}
\hspace{2cm}
\includegraphics[width=.35\textwidth]{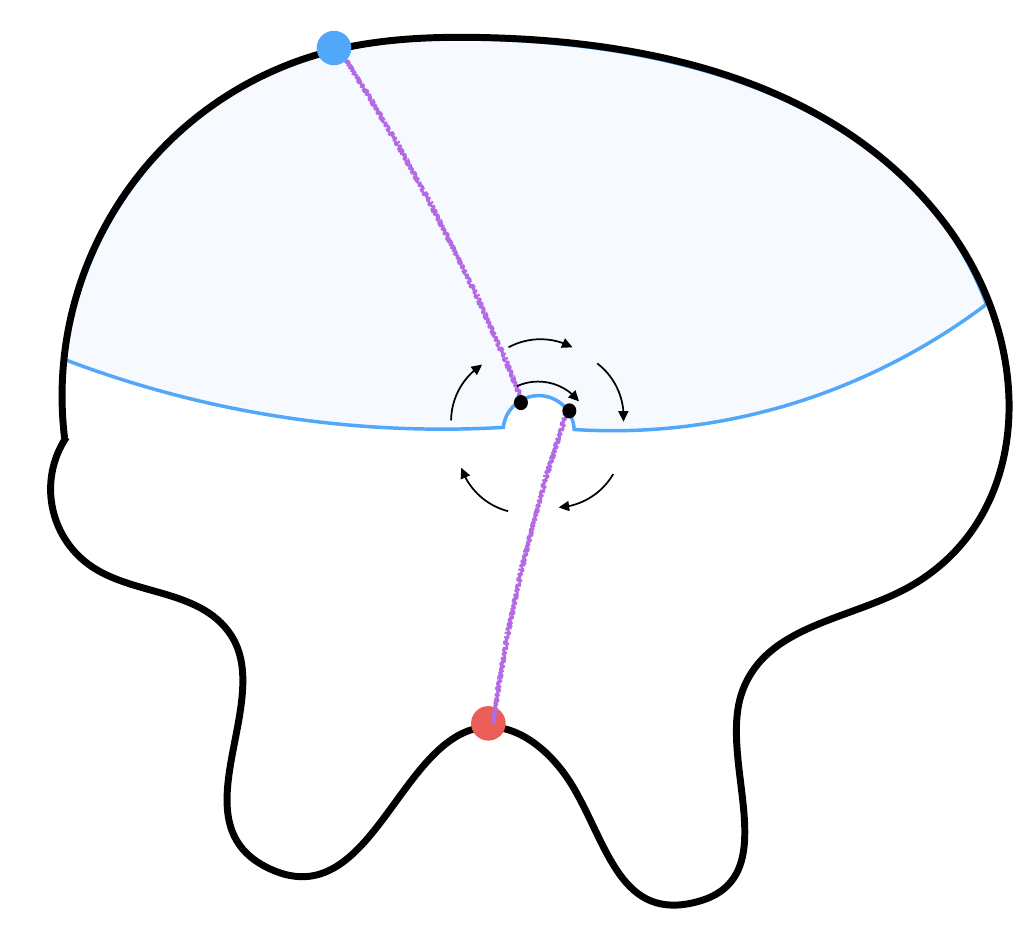}
\caption{\label{fig:cart} Modular flow in Euclidean is described via an integral over three correlation functions across the boundaries of the three regions above. The shaded regions are determined by the known Green's function while the unshaded region involves the modular flow correlator which in general is not known. It is however constrained close to the RT surface which is in the center of the figure. In the right figure we have done one of the boundary integrals assuming the approximation to local flow using the vector field $\chi_E$ which is also shown here.
}
\end{figure}

Plugging in \eqref{midterm} and \eqref{outside} into the master integrals \eqref{comp} and taking the limit $n \rightarrow 1$ will be our starting point for making approximations. A cartoon of this result is in Figure~\ref{fig:cart}.  
Assuming dominance for the $Z_i$ integrals near the entangling surface we can replace:
\be
\label{continue}
{\rm Tr}_a \rho_{(1)}^{1-is} \phi(Z_1) \rho_{(1)}^{is} \phi(Z_2) \approx 
G(Z_1, e^{ s \chi }(Z_2) )  + \ldots
\ee
where $G=G_{n=1}$ and $\chi$ is a vector field that generates boosts in real time close to the entangling surface and away from the entangling surface we choose it arbitrarily. This vector was introduced already in \eqref{chi}. For now we will mostly be interested in Euclidean modular flow
so we have wick rotated $s = -  i \theta$ and $\chi_E = - i \chi$ where $\chi_E$ now generates rotations around the entangling surface. Picking gaussian normal coordinates:
\be
ds^2 = dw d\bar{w} + h_{ij}(y) dy^i dy^j + \ldots
\ee
this vector can be chosen to be:
\be
\chi_E = i \left( w \partial_w - \bar{w} \partial_{\bar w} \right)
\ee
such that $e^{ \theta \chi_E}(w) = e^{ i 2\pi \theta} w$. Extending $\chi_E$ aribtrarily away from the entangling surface we will make the replacement \eqref{continue} in \eqref{comp} everywhere even far away from the entangling surface. This will be justified later when we show the dominate contribution comes from close to the entangling surface. We find:
\begin{align}
g_{n=1,\theta}(x,y) \approx \int_{\partial S_1 } \int_{\partial S_2 }    K(x,Z_1) n_1\cdot \left.\mathop{\nabla}^\leftrightarrow\right._{Z_1} G(Z_1,e^{\theta \chi_E}(Z_2))  \times n_2\cdot \left.\mathop{\nabla}^\leftrightarrow\right._{Z_2} K(Z_2,y)
\end{align}
At this point it is important that we pick the regions $S_1$ and $S_2$ to not be overlapping on the single remaining replica. 
Then we can undo $Z_1$ integral, as along as  $e^{\theta \chi_E}(Z_2)$ does not come inside the region $S_1$. Such that:
\be
\label{stillint}
g_{1,\theta}(x,y) \approx \int_{\partial S_2}  K_n(x,e^{i \theta \chi_E}(Z_2) )   n_2\cdot \left.\mathop{\nabla}^\leftrightarrow\right._{Z_2} K_n(Z_2,y)
\ee
We can evaluate this in the saddle approximation for  the integral over $\partial S_2$ after
replacing the $K's$ with their semi-classical approximations for large $m$. See the right part of Figure~\ref{fig:cart}. We pick $\partial S_2$
to have a small segment that wraps a semi-circle around the entangling surface at some fixed distance $\rho$. Self consistency of the approximation demands that there should be a saddle along this small segment where $\rho$ can be taken small. If $\rho$ is indeed small then we can approximate the saddle by expanding around geodesics that go to the entangling surface at $\xi$. The lengths of the sum of such geodesics is:
\begin{align} \nonumber
\ell(x,X_A(\xi))  + \ell(y,X_A(\xi)) + &\frac{1}{2} \left( e_x^{-1} (e^{-i 2\pi \theta} \partial_\tau w_x  \delta \bar{w} 
+e^{i 2\pi \theta} \partial_\tau \bar{w}_x  \delta w)  
+ e_y^{-1} ( \partial_\tau w_y  \delta \bar{w} 
+\partial_\tau \bar{w}_y  \delta w)    \right)  \\
&  \qquad \qquad 2 h_{ij}(\xi) \delta y^i (e_x^{-1} \partial_\tau y^j_x +e_y^{-1} \partial_\tau y^j_y  )
\label{fterms}
\end{align}
where the deformations away from the entangling surface are given, in holomorphic coordinates, by $\delta w$ for the $x$-geodesic
and $e^{-i 2\pi \theta} \delta w$ for the $y$-geodesic. This is the jump condition that is implied by the
expression in \eqref{stillint}. We must constrain $\delta w = \rho e^{ i \phi}$ for fixed $\rho$. Minimizing over $\delta y^i$
and $\phi$ give Wick rotated versions of the conditions in (\ref{match1}-\ref{match2}). Note in particular these equations say nothing about $\phi$. So there is a family of piecewise geodesics that are parameterized by $\tau$, satisfy the jump condition at the $\rho$ surface:
\be
w_x = e^{ - i 2\pi \theta} w_y
\ee
and to leading order in $\rho$ satisfy the momentum conditions of \eqref{match1}. There is one value of $\theta$ for which the geodesic exactly passes through the entangling surface, giving Rule $1$, and as long as the minimal value of $\rho$ that still allows for a saddle can be taken small these other cases give rule $2$ with  $\mathcal{O}(\rho_{\rm min}^2)$ corrections to the actually length of the geodesic away from $\ell(x,X_A(\xi))  + \ell(y,X_A(\xi))$. Note that the computation that we have outlined is the natural Wick rotated version of Rule $2$. Indeed we get the version of Rule 2 that is convenient for calculating with, that involves a discontinuous geodesic across the null surface $\mathcal{H}^+$. So after passing to real times, by simultaneously deforming the regions $S_2$ to the real time section and then pushing it to $\partial \mathcal{E}_a$, and making the modular flow parameter real this gives a derivation of this rule. Note that the wick rotation here sends the euclidean Green's function to the Feynman ordered Green's function.

One may worry that while this is one possible saddle for the integrals in \eqref{comp}, there might be others that are outside of the regime of the approximation used in \eqref{continue}. And these may dominate. If we can't calculate in this case it seems hard to check this cannot happen. We will now give strong evidence that 
that we have found the dominant saddle at large $m$, once we impose the matching condition on the parameters $x,y,s$. This argument works in real times where we assume the underlying Euclidean manifold has some time reflection symmetry.

Let us first consider the case where the two points on opposite wedges and take $s$ to be real. Consider the Cauchy-Schwarz bound:
\be
\label{csbound}
|\left< \psi \right| \mathcal{O}(x) \Delta^{is}_A  \mathcal{O}(y) \left|\psi \right>|^2
\leq \left< \psi \right| \mathcal{O}(x) \Delta^{1/2}_A  \mathcal{O}(x) \left| \psi \right>
\left< \psi \right| \mathcal{O}(y) \Delta^{-1/2}_A  \mathcal{O}(y) \left| \psi \right>
\ee
Since the $y$ operator is inserted in $\bar{A}$ we have the relation $J_A \Delta^{-1/2}_A  \mathcal{O}(y) \left| \psi \right> = \mathcal{O}(y)^\dagger \left| \psi \right> =   \mathcal{O}(y)\left| \psi \right> $.
So indeed both of the correlators on the right hand side of \eqref{csbound} are mirror operator correlators. They are thus both computable in the geodesic approximation. Indeed we have exactly computed the right hand side using a different replica trick for $\Delta^{1/2}$ in Section~\ref{dhalf}. If we consider parameters such that $\partial_\tau y^j = 0$ at the RT surface for the geodesic on the left hand side of \eqref{csbound} then it is not hard to see that the classical lengths are the same for both sides of the bound. Thus Cauchy-Schwarz is approximately saturated. It is not exactly saturated because we have not computed the one loop determinants that will differ by an order $1$ amount. 
While this fact is intriguing to us, we will leave exploration of it to future work. We only need to know that if any other saddle exists for the left hand side of \eqref{csbound}, it must have longer length or the same length - because this saddle is already saturating an upper bound.

We can make a similar argument when the operators are in the same wedges with:
\be
\label{csbound2}
|\left< \psi \right| \mathcal{O}(x) \Delta^{is+1/2}_A  \left| \mathcal{O}(y) \right>|^2
\leq \left< \psi \right| \mathcal{O}(x) \Delta^{1/2}_A  \mathcal{O}(x) \left| \psi \right>
\left< \psi \right| \mathcal{O}(y) \Delta^{1/2}_A  \mathcal{O}(y) \left| \psi \right>
\ee
For most of the results of this paper we use the case where the momentum $\partial_\tau y^j$ at the entangling surface is zero or small, so this is all the evidence we really need. 
In other cases where the momentum along the surface is non-zero one is still fairly close to saturating this bound, and we think this is strong evidence that no other saddles may dominate. 

We note that the special curves that do saturate the Cauchy-Schwarz bound seem to play a special role. They can be explicitly written as:
\be
y^\perp_x(s) \equiv y_{(x,\xi(x))}(s)
\ee
It would be interesting to understand better the meaning of these special curves, and weather they can be used to more generally constrain modular flow.

\section{Discussion}

In this paper we have given new rules for computing modular flow of heavy probe operators in AdS/CFT. These rules can be used to extract causal properties of the bulk theory such as entanglement wedge nesting. We also highlighted the importance of the so called mirror operators who's correlation function with the original operator are computed by reflected geodesics. Such mirror operators allow one to map out the entangling surface from a boundary region $A$ with a map $\xi(x)$. 

The rules we have introduced here can be used to extract many other properties of the bulk. 
One might eventually hope to be able to write Einstein's equations, or perhaps some causal property directly related to Einstein's equations such as the Quantum Focusing Condition in terms of these correlators. We leave such extensions to future work.

\acknowledgments

We would like to thank Aitor Lewkowycz, Xiao Liang Qi, Xi Dong and Mukund Rangamani for discussions on their work and for sharing an early draft of their paper with us. This work was supported by the Department of Energy contract DE-SC0015655 and SC0019183
\newpage
\appendix
\allowdisplaybreaks
\section*{Appendix} 
\section{Re-deriving the QNEC results}
\label{app:qnec}

In this section, we focus on the class of correlators studied in \cite{Balakrishnan:2017bjg}
\begin{eqnarray}
f(s)=\frac{\la\psi|\OO(x_B)\Delta_B^{is}\Delta_A^{-is}\OO(x_{\bar{A}})|\psi\ra}{\la\Omega|\OO(x_B)\Delta_{B,\Omega}^{is}\Delta_{A,\Omega}^{-is}\OO(x_{\bar{A}})|\Omega\ra}
\end{eqnarray}
here $A$ and $B$ are two sub-regions on the boundary, which satisfied $\mathcal D(B)\subset\mathcal D(A)$. The entangling surfaces $\Sigma_{A,B}$ intersect with boundary at $\partial A$ and $\partial B$ and these two entangling surfaces are light like separated there.  For simplification, set $\partial A$ and $\partial B$  to be symmetric with respect to the origin (of our null coordinates) and also take the operator insertions $x_B,x_A$ symmetric about this surface.   $\Sigma_{A,B}$ have two local normal null coordinates $k_{\pm}$ satisfying (\ref{1.1.1}). We work in FG coordinates (\ref{3.3.1}).

According to \textit{rule 2}, the geodesic is separated into three parts: left, middle and right and discontinuous along the null light-sheets that pass through $\Sigma_{A,B}$ in the $k_+^{A,B}$ direction. By symmetry we only need to consider the left (L) and middle (M) segments and we can parameterize this with $-1< \tau < 0$. We will  find that at the mid point, and to leading order in the state, $x^\pm_M(0) =0$ and $x^i_M(0) = \xi^i_{\rm min}$ for some as yet unfixed $ \xi^i_{\rm min}$. Note that $\xi^i = (\vec{y},z)$.  
The other boundary conditions are:
\begin{eqnarray}\label{1.20}
&&x_L(-1)=x_B \\
&&  x_L^i(-1/2)=\xi_B^i,\quad x_L^-(-1/2)=X^-_{B}(\xi_B),\quad x_L^+(-1/2)=v^*_B\\\label{1.21}
&&x_M^i(-1/2) =\xi_B^i,\quad x_M^-(-1/2)=X^-_{B}(\xi_B),\quad x_M^+(-1/2)=v^*_B e^{-s}
\end{eqnarray}
where $\tau=-1/2$ is the matching point between the left and middle segments and varying over $v^*_B$ and $\xi_B$ gives the other set of matching condition for the momentum of the geodesic as it crosses the ``horizons'' of RT surfaces: (\ref{match1}) and (\ref{match2}).


\subsection{Example: geodesic in pure AdS}

In this parts, we would solve geodesics between $x_{\bar A}$ and $x_B$ in vacuum with double modular flow transformation.  Of course in vacuum these are simple boosts and the result should be computed by the length of a single geodesic after the action of these boosts. However it is useful to first construct the relevant piecewise continuous geodesic in this case - we will then use this to perturb around vacuum.

We can solve for this geodesic by imposing boundary conditions (\ref{1.20}) and matching conditions (\ref{match1}) and (\ref{match2}) in AdS.  We choose to parameterize the curve by $\tau \rightarrow u$ which is a gauge choice. We proceed as follows.

A general geodesics between two boundary points are half circles in the bulk,
\begin{eqnarray}\label{1.22}
z_{vac}^2(u)=z_{min}^2+K(u-u_{min})^2,\qquad K=-\frac{\Delta u}{\Delta v}
\end{eqnarray}
$K$ is the slope of the geodesic projection onto the $z$ plane, which is a constant in pure AdS.  $\Delta u(v)$ is the distance between two boundary points.  $z_{\rm min}$ is the lowest points of geodesic.  In pure AdS, we have $z_{\rm min}^2=-\frac{\Delta u\Delta v}{4}$.\par

\begin{figure}[h!]
	\centering
	\includegraphics[width=0.9\linewidth]{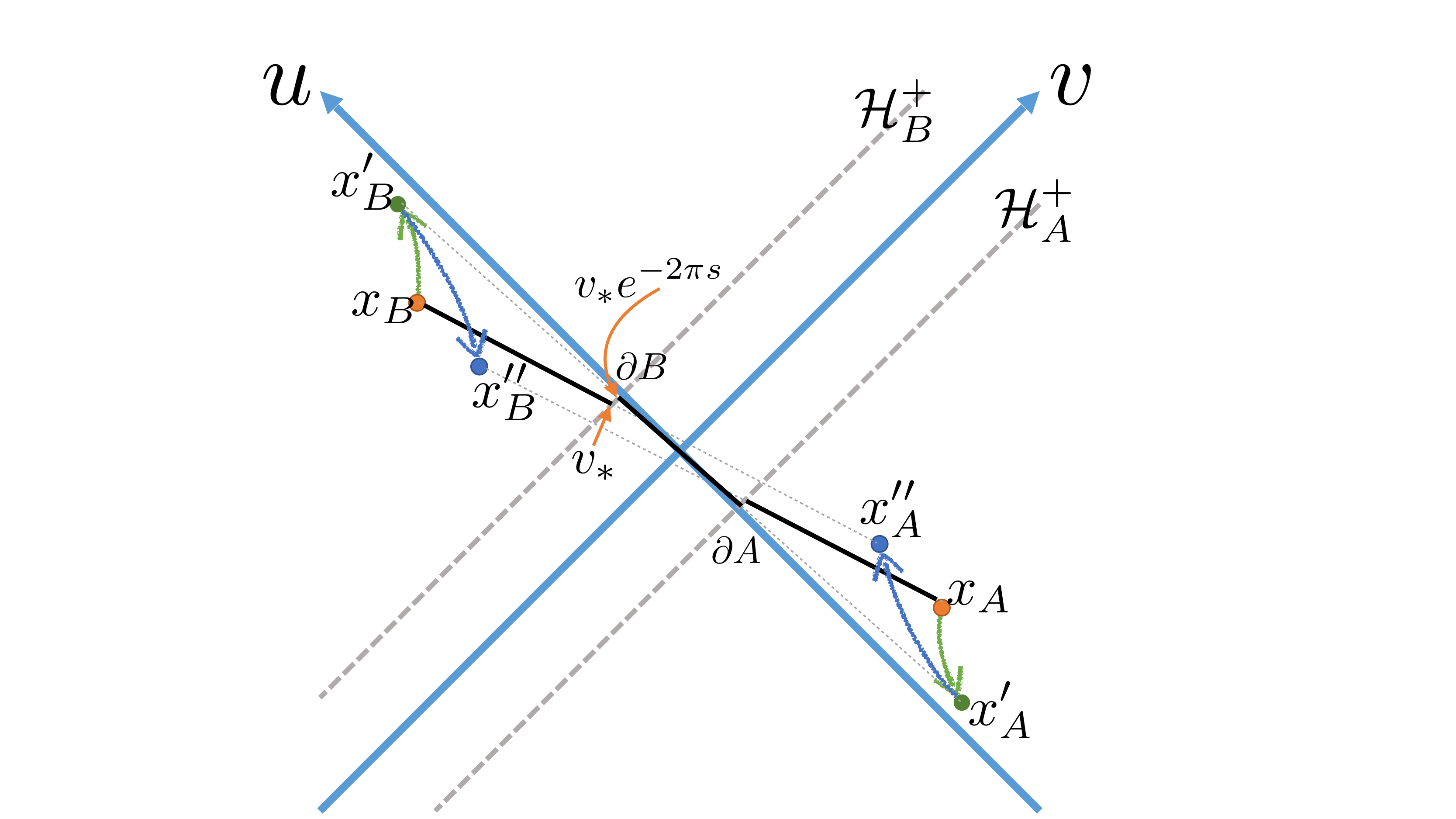}
	\caption{Operators on the boundary transform under modular flow in various ways. These result in the various geodesics that make up the piecewise segments of the flowed geodesic.}
	\label{fig:fig2}
\end{figure}

We will use this basic  geodesic to solve for the piecewise continuous geodesic satisfying \eqref{1.20}: by finding the boundary points that each geodesic segment is anchored to.  The left part of geodesics is anchored between two local operators, one is $\OO(x_B)$.  Then it is not hard to guess that one can find the other by transforming the operator $\OO_{\bar{A}}$ under double modular flows.  Therefore, we would get a local operator at another point defined by $\OO(x_{\bar{A}}'')$,
\begin{eqnarray}
\OO_{\bar{A}}(x_A'')=\Delta_B^{is}\Delta_B^{-is} \OO_{\bar{A}} \Delta_A^{is}\Delta_A^{-is}
\end{eqnarray}
acting with two boosts in sequence results in a transformation,
\begin{eqnarray}
x_{\bar{A}}''=(u_{\bar{A}}'',v_{\bar{A}}'')=(-u_B+\delta x^-_s,-v_B),\qquad \delta x^-_s=\delta x^-(1-e^{-2\pi s})
\end{eqnarray}
we can find that the left part of geodesic parameterized by $u$ reads
\begin{eqnarray}
&&v_L(u)=\frac{\Delta v}{\Delta u_s}\big(u-\frac{\delta x^-_s}{2}\big),\qquad \Delta v=v_B-v_A\\
&&z_L(u)=\sqrt{\bigg(-\frac{\Delta v}{\Delta u_s}\bigg)\bigg(\frac{\Delta u_s}{2}-u+\frac{1}{2}\delta x_s^-\bigg)\bigg(\frac{\Delta u_s}{2}+u-\frac{1}{2}\delta x_s^-\bigg)},\quad\\
&& e_L(u)=-\frac{\Delta v}{2 z_{\rm min}^2}
\end{eqnarray}
where $e^2 =  \partial_\tau x \cdot \partial_\tau x$ is the ``einbein'' and $\Delta u_s=\Delta u-\delta x_s^-$.

We can find the middle part of geodesic in the same way.  The two boundary points of middle geodesics can be found as:
\begin{eqnarray}
&&\OO_{B}(x_B')=\Delta_B^{-is} \OO_{B} \Delta_B^{is},\qquad \OO_{\bar{A}}(x_A')=\Delta_A^{-is}\OO_{\bar{A}} \Delta_A^{is},\\ \nonumber
&&x_B'(u_B',v_B')=x_B'\big(e^s(u_B-\frac{\delta x^-_s}{2}),v_Be^{-s}\big)\, \quad x_{\bar{A}}'(u_{\bar{A}}',v_{\bar{A}}')=x_{\bar{A}}'\big(e^s(u_{\bar{A}}+\frac{\delta x^-_s}{2}),v_{\bar{A}}e^{-s}\big),
\end{eqnarray}
For which the middle part of geodesics is:
\begin{eqnarray}
&&v_M(u)=e^{-4\pi s}\frac{\Delta v}{\Delta u_s}u\\
&&z_M(u)=\sqrt{\bigg(-\frac{\Delta v}{\Delta u_s}\bigg)\bigg(\frac{\Delta u_s}{2}-e^{-s}u\bigg)\bigg(\frac{\Delta u_s}{2}+e^{-s}u\bigg)}\\
&& e_M(u)=-e^{-2\pi s}\frac{\Delta v}{2 z_{\rm min}^2}
\end{eqnarray}
Note that $K_M=\big(-\frac{p_v}{p_u}\big)_M=-e^{-4\pi s}\frac{\Delta v}{\Delta u_s}=e^{4\pi s}K_L$, which satisfied the boost matching conditions (\ref{match1}) and (\ref{match2}). Also note the relation between the ``einbeins'': $e_M = e^{-2\pi s} e_L$ at the matching point.

\subsection{General state}
In some general states of $\psi$, the metric near boundary has a Fefferman-Graham expansion (\ref{3.3.1}).
Two RT surfaces have expansion near the boundary (\ref{3.3.2}).
Expand the correlator $f(s)$ at first order of $G_N$, we have $f(s)\simeq 1-\Delta_{\OO}(\ell-\ell^0)+\OO(G_N^2)$, where $\ell^0$ is the geodesic length in vacuum AdS. To find the change in length we simply perturb the geodesics we found in the previous subsection by adding the appropriate metric fluctuation:
\begin{eqnarray}
&&f(s)=1- \frac{\Delta_{\OO}}{2}\Bigg(\int_{-1}^1 d\tau\, \frac{\delta g_{\mu\nu}^{\psi}\partial_{\tau}X^{\mu}\partial_{\tau}X^{\nu}}{e(\tau)}+\bigg(\delta X^{\mu}\frac{g_{\mu\nu}^{0}\partial_{\tau}X^{\nu}}{e(\tau)}\bigg)\Lvert^{1/2}_{-1/2} \Bigg)+ \ldots \\
&&	\delta g_{\mu\nu}^{\psi}=\frac{16\pi G_N}{d}z^d\la T_{\mu\nu}\ra_{\psi}
\end{eqnarray}
where we have used the symmetry of the undeformed geodesic to extend this in the natural way to the right segment with $0 < \tau < 1$. 

Note that  the $\delta X^{\mu}$ part accounts for the change of geodesic length due to change of the end points at $\mathcal{H}^+_{A,B}$. Since the unperturbed geodesic is effectively off-shell at the matching points, variations of the geodesic can lead to a boundary term. Locally, if the entangling surface moved in the $v$ direction by $\delta X^+$, then the end points of both the middle and left part of geodesic would lead to a combined changed of $\delta X^+(1-e^{-2\pi s})$ due to the matching condition on the horizon of entangling surface (\ref{1.21}). Note that this term clearly vanishes when $s=0$. 

Parameterizing the geodesic by $u$ and estimating the geodesics with the vacuum AdS result we have: 
\begin{eqnarray}\label{1.23}
f(s)
&=&1+\frac{16\pi G_N\Delta_{\OO}}{d}\frac{1}{\Delta v}\bigg(-\frac{\Delta v}{\Delta u_s}\bigg)^{\frac{d}{2}}\Bigg(\int^{u_B}_{\frac{\delta x^-}{2}}du \la T_{uu}\ra_{\psi}b_L(u)+\int_{-\frac{\delta x^-}{2}}^{\frac{\delta x^-}{2}}du \la T_{uu}\ra_{\psi}b_M(u)\non\\
&&\qquad+\int^{\frac{\delta x^-}{2}}_{u_{\bar{A}}}du \la T_{uu}\ra_{\psi}b_R(u)+ \frac{1-e^{-2\pi s}}{2\pi}\bigg(b_M(\frac{\delta x^-}{2})\mathcal{P}_-^A-b_M(-\frac{\delta x^-}{2})\mathcal{P}_-^B\bigg)\Bigg)\non\\
\end{eqnarray}
here we have
\begin{eqnarray}
b_M(u)&=&e^{2\pi s}\bigg(\frac{\Delta u_s}{2}-e^{-2\pi s}u\bigg)^{\frac{d}{2}}\bigg(\frac{\Delta u_s}{2}+e^{-2\pi s}u\bigg)^{\frac{d}{2}}\non\\
b_L(u)&=&\bigg(\frac{\Delta u_s}{2}-u+\frac{1}{2}\delta x_s^-\bigg)^{\frac{d}{2}}\bigg(\frac{\Delta u_s}{2}+u-\frac{1}{2}\delta x_s^-\bigg)^{\frac{d}{2}}\non\\
b_R(u)&=&\bigg(\frac{\Delta u_s}{2}-u-\frac{1}{2}\delta x_s^-\bigg)^{\frac{d}{2}}\bigg(\frac{\Delta u_s}{2}+u+\frac{1}{2}\delta x_s^-\bigg)^{\frac{d}{2}}\,.
\end{eqnarray}
Note that $b_M(\lambda_-)=e^{2\pi s}b_L(\lambda_-)$, $b_M(-\lambda_-)=e^{2\pi s}b_R(-\lambda_-)$, which satisfied the local matching condition.  These results are exactly the same as the results in \cite{Balakrishnan:2017bjg} after shifting the $u$ coordinate in the paper $u\rightarrow u-\delta x^-/2$ and $\Delta u\rightarrow \Delta u+\delta x^-$. Of course the methods were very different. 

\bibliography{mod-toolkit}



\end{document}